\documentclass{aa}

\usepackage[varg]{txfonts}
\usepackage{graphicx}
\usepackage{microtype}
\usepackage[english]{babel}
\usepackage{slashbox}
\usepackage{colortbl}

\usepackage{natbib}
\bibpunct{(}{)}{;}{a}{}{,} 

\graphicspath{{figures/}}

\let\corr\relax  
\let\mean\relax  
\let\rms\relax  
\DeclareMathOperator*{\corr}{corr}
\DeclareMathOperator*{\mean}{mean}
\DeclareMathOperator*{\rms}{rms}
\newcommand{\FWHM}{$\mathrm{FWHM}$}

\newcommand{\mps}{m\,s$^{-1}$}
\newcommand{\dqinv}{\delta q_{\alpha}^{\mathrm{inv}}}
\newcommand{\akern}{\mathcal{K}^{\alpha}_{\beta}}
\newcommand{\lab}{\tens{\Lambda}^{ab}}
\newcommand{\w}{w^{\alpha}}
\newcommand{\dq}{\delta q^{\beta}}
\newcommand{\ssp}{sound-speed perturbations}

\newcommand{\dif}[1]{\mathrm{d}{#1}\,}
\newcommand{\ddif}[1]{\mathrm{d}^2\vec{{#1}}\,}
\newcommand\norm[1]{\left\lVert#1\right\rVert}

\newcommand{\ocfigure}[4][tbp]{
    \begin{figure}[#1]
        \resizebox{\hsize}{!}{\includegraphics{#2}}
        \caption{#3}
        \label{#4}
    \end{figure}
}

\newcommand{\tcfigure}[4][tbp]{
    \begin{figure*}[#1]
        \centering
        \includegraphics[width=17cm]{#2}
        \caption{#3}
        \label{#4}
    \end{figure*}
}

\newcommand{\wfigure}[4][tbp]{
    \begin{figure*}[#1]
        \sidecaption
        \includegraphics[width=12cm]{#2}
        \caption{#3}
        \label{#4}
    \end{figure*}
}

\usepackage{color}
\usepackage[normalem]{ulem}
\usepackage{enumitem}

\begin{document}

\title{
    One-sided arc averaging geometries in\\time--distance local helioseismology
    }

\author{
    David Korda\inst{1,}\inst{2}
    \and
    Michal {\v S}vanda\inst{2,}\inst{3}
    \and
    Thierry Roudier\inst{4}
}

\offprints{David Korda, \\ \email{korda@sirrah.troja.mff.cuni.cz}}

\institute{
    Department of Geosciences and Geography, Faculty of Science, University of Helsinki, Gustaf H\"{a}llstr\"{o}min katu 2, FI-00014, Helsinki, Finland
    \and
    Astronomical Institute of Charles University, Faculty of Mathematics and Physics, V~Hole\v{s}ovi\v{c}k\'ach 2, CZ-18000, Prague 8, Czech Republic
    \and
    Astronomical Institute of Czech Academy of Sciences, Fri\v{c}ova 298, CZ-25165, Ond\v{r}ejov, Czech Republic
    \and
    Institut de Recherche en Astrophysique et Plan\'etologie, Universit\'e de Toulouse, CNRS, UPS, CNES 14 avenue Edouard Belin, F-31400, Toulouse, France
} 

\abstract
{The study of solar oscillations (helioseismology) has been a very successful method of researching the Sun. Helioseismology teaches us about the structure and mean properties of the Sun. Together with mid-resolution data, the local properties were uncovered in quiet-Sun regions. However, magnetic fields affect the oscillations and prevent us from studying the properties of magnetically active regions with helioseismology.}
{We aim to create a new methodology to suppress the negative effects of magnetic fields on solar oscillations and measure plasma properties close to active regions.}
{The methodology consists of new averaging geometries, a non-linear approach to travel-time measurements, and a consistent inversion method that combines plasma flows and \ssp{}.}
{We constructed the one-sided arc averaging geometries and applied them to the non-linear approach of travel-time measurements. Using the one-sided arc travel times, we reconstructed the annulus travel times in a quiet-Sun region. We tested the methodology against the validated helioseismic inversion pipeline. We applied the new methodology for an inversion for surface horizontal flows in a region with a circular H-type sunspot. The inverted surface horizontal flows are comparable with the output of the coherent structure tracking, which is not strongly affected by the presence of the magnetic field. We show that the new methodology suppresses the negative effects of magnetic fields up to outer penumbra. We measure divergent flows with properties comparable to the moat flow.}
{The new methodology can teach us about the depth structure of active regions and physical conditions that contribute to the evolution of the active regions.}

\keywords{
    Sun: helioseismology -- Sun: oscillations -- Sun: interior -- Sun: sunspots
}

\authorrunning{D. Korda et al.}
\maketitle


\section{Introduction}

Helioseismology studies the solar interior via properties of the solar oscillations caused by the sound and the surface gravity waves that propagate through the Sun. The waves are affected by plasma conditions along their trajectory. Typical methods of local helioseismology utilised for inferring the plasma properties in mid resolution are the ring-diagram method \citep{Hill_1988} and the time--distance method \citep{Duvall_1993}. The ring-diagram method is based on the direct study of an average power spectrum and its resonant frequencies. The resonant frequencies are also sensitive to the plasma flows and the sound speed via the Doppler effect. Time--distance helioseismology estimates the properties of the plasma flows and the \ssp{} via the travel times of the waves between two points on the solar surface. The travel times are naturally affected by the plasma flows and sound speed. These methods have been utilised in many studies \citep[e.g.][]{Basu_1999, Kosovichev_2000, Bruggen_2000, Haber_2002, Zhao_2004, Couvidat_2006, Zaatri_2006, Hindman_2006, Komm_2007, Komm_2008, Bogart_2008, Hindman_2009, Zhao_2010, Baldner_2011, Svanda_2013}.

Achievements of local helioseismology in quiet-Sun regions led researchers to take advantage of its power in active regions. However, the inverted plasma flows and \ssp{} were usually very chaotic \citep{Zhao_2001, Zhao_2003, Zhao_2010}. The inconsistencies were explained as an effect of magnetic fields. The amplitudes of the measured travel times of the waves are modified by the presence of the magnetic field \citep{Cally_2007}. The changes were misinterpreted as the plasma flows or the \ssp{}. Furthermore, different helioseismic methods returned different results, which were mutually inconsistent \citep{Gizon_2009}.

The active regions have been intensively studied using numerical simulations \citep[e.g.][]{Rempel_2009, Rempel_2011}. The realistic numerical simulations have allowed the study of how the strong magnetic field affects the helioseismic products \citep{Moradi_2009, Braun_2012, DeGrave_2014, Braun_2019}. To avoid issues with the magnetic fields, researchers have tried to mask out the active regions \citep[e.g.][]{Liang_2015a}. \citet{Zhao_2003} inverted for plasma flows and \ssp{} beneath a sunspot. They have also compared results with and without a masked-out active region but did not find significant differences between the corresponding results. \citet{Korzennik_2006} inverted for a set of models of \ssp{}. For each model, he masked out a circular region around a sunspot with different radius and found out that `plume' \ssp{} below an active region are probably caused by the surface contamination.

A different strategy was chosen by \citet{Duvall_2018}. They observed points far from a sunspot and selected waves with two skips, where the first skip is inside the sunspot. \citet{Hughes_2005} compared the inverted \ssp{} using single-skip and double-skip geometries. They found a good agreement at depths larger than about 5~Mm below a sunspot. The waves with two skips cannot be separated from those with one skip if the observed points were closer to each other. For this reason, they distinguished a lack of waves propagating through the shallow layers in the double-skip geometry. For this reason, this type of measurement is not suitable in shallow layers.

Taking the arguments given above into account, the sub-surface structure and the dynamics of the active regions are still unknown. This is in contrast with the fact that the large active regions are potentially dangerous for a technical civilisation via solar flares.


\section{Motivation}

Around an evolved sunspot, divergent horizontal flows called moat flows are observed. The amplitude of a moat flow is about 500~\mps{} (see e.g. \citet{Svanda_2014}). Numerical simulations \citep{Rempel_2009} predicted a uniform outflow from the centre of the sunspot. The helioseismic inversions \citep[e.g.][]{Zhao_2010} resulted in a multi-layer structure, in which the direction of the flows and the sign of the \ssp\ changed repeatedly{}. Moreover, helioseismic results depend on the specific approach \citep{Gizon_2009}. Therefore, the depth profile of the moat flow is unknown unless it has a large amplitude.

In the presence of the magnetic field, the conversion between sound waves and magnetoacoustic waves appears. The magnetoacoustic waves are not trapped inside the Sun and escape into the atmosphere \citep{Schunker_2013}. In this case, we can observe the wave at only one of the points between which we measure the travel times. Therefore, the amplitudes of the detected waves are lower in active regions. This effect is usually misinterpreted as plasma flows or \ssp{}. The amplitude of the detected waves (the amplitude of the cross-correlation function $C$ defined in Sect.~\ref{sec:tt}) explicitly appears in the linear approach of the travel-time measurements \citep[GB04;][]{GB04}, while it does not enter non-linear approach \citep[GB02;][]{GB02}. Therefore, the GB02 travel times are more suitable for studies of the active regions. The comparison between GB02 and GB04 travel times in an active region was done by \citet{DeGrave_2014} who showed the advantages of the GB02 approach in active regions.

Time--distance helioseismology uses the averaging geometries. These increase the signal-to-noise ratio. The most often used averaging geometries assume quasi-uniform properties of the observed data since the cross-correlation is computed between the observed data at a point and the data averaged over the annulus or quadrant around the point \citep[e.g.][]{Duvall_1997, Kosovichev_2000, Svanda_2013}. Then, the final travel times are computed from a combination of the individual travel-time measurements. For example, in quadrant geometry the travel times sensitive to flows in the longitudinal direction are computed as the difference between the travel times measured in and against the direction of the solar rotation. This effectively makes the final averaging geometries (and the travel times) sensitive to the conditions all around the observed point. Therefore, these averaging geometries are very useful in quiet-Sun regions, but not in the vicinity of active regions, because they average the active and the quiet regions together. An effect of an active region to the cross-correlation function was demonstrated, for example, by \citet{Cameron_2008}. They computed the cross-correlation function and showed that an active region absorbs non-magnetic waves (see their Fig.~4). Similar results were obtained by \citet{Gizon_2009}, who measured the absorption of surface gravity waves and sound waves with different radial orders. The absorption coefficients were between 49\% and 57\%. Moreover, the two-sided character of the averaging geometries causes omnidirectional disturbances even in quiet-Sun regions close to an active region, as was pointed out by \citet{Hughes_2005}.

\ocfigure{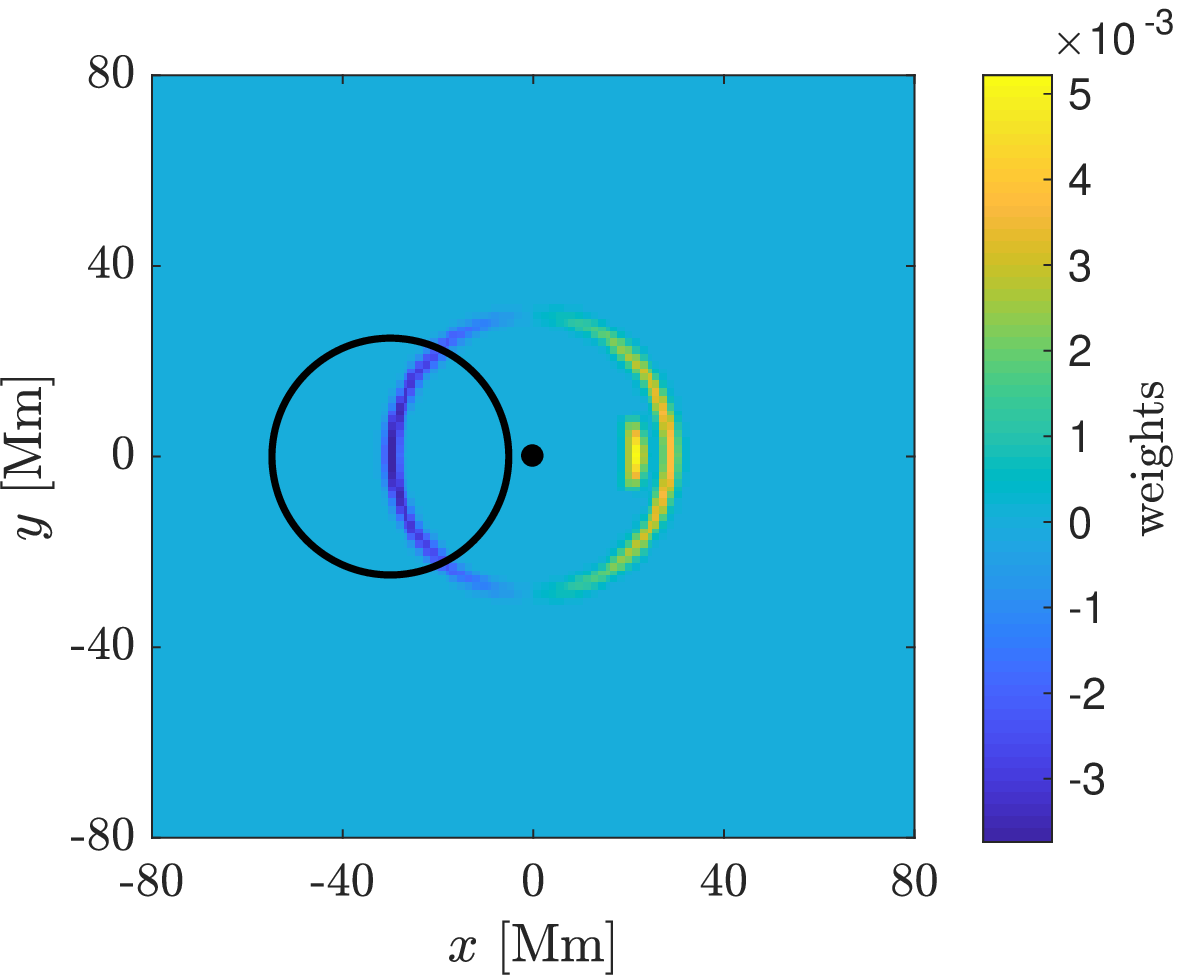}
{Examples of the annulus (the outer annulus) and the arc (the inner arc) averaging geometries. The black dot is the point where the travel time is measured. The colours correspond to the weights of the averaging geometries. The black circle delimits the artificial active region. The annulus geometry averages the quiet and the active regions because of its two-sided character; the observed point in the one-sided arc geometry is not affected by the active region. For display purposes, the radii of the averaging geometries differ, the full width at half maxima (\FWHM{}s) of the Gaussian annuli were enlarged, and the weights of the arc geometry were divided by 12.}
{fig:ave_geoms}

As mentioned in the introduction, the possible way to correctly measure plasma parameters below a sunspot is to utilise the double-skip geometry. As a trade-off, the user cannot reliably invert for the parameters in the first few megametres \citep{Hughes_2005}. The way to avoid these issues is to abandon the approaches of the double-skip geometry and the two-sided averaging geometries.

In this paper, we introduce a new type of averaging geometry: one-sided arc geometries. As seen in Fig.~\ref{fig:ave_geoms}, the one-sided geometries do not average the quiet and the active regions together. Naturally, the one-sided averaging geometries do not need to use the double-skip geometry. Therefore, the one-sided averaging geometries can be effective in the vicinity of the active regions, where the moat flow appears. In Fig.~\ref{fig:ave_geoms}, we plotted examples of the annulus (the outer annulus) and the arc (the inner arc) averaging geometries. In this case, both geometries are sensitive to flows in the $x$ direction. The black dot at the coordinates $\left[0, 0\right]$ is the point at which the travel time is measured, the colours indicate the weights of the averaging geometries, and the black circle corresponds to the edge of an artificial active region. In the case of the annulus geometries, the central point is affected by the active region, even though the point itself is not inside the active region (part of the annulus is inside it). This is not the case of the one-sided arc geometries. Therefore, we can obtain unspoiled information about a much closer vicinity of the active region if we combine the one-sided arc geometries and the non-linear GB02 travel-time approach.


\section{Travel times}
\label{sec:tt}

The travel times $\tau$ of the sound and the surface gravity waves, used in the time--distance helioseismology, are measured between the two points $\vec{r}_1$ and $\vec{r}_2$ on the solar surface. The travel time $\tau \left(\vec{r}_1, \vec{r}_2 \right)$ is defined as the time lag that maximises the temporal cross-correlation $C \left( \vec{r}_1, \vec{r}_2, t \right)$ of the observed signal $\psi\left(\vec{r}, t\right)$ at the points $\vec{r}_1$ and $\vec{r}_2$ \citep{GB02}:
\begin{align}
    C \left( \vec{r}_1, \vec{r}_2, t \right) &= \frac{1}{T} \int \limits_{-\infty}^{\infty} \dif{t'} \psi \left( \vec{r}_1, t' \right) \psi \left( \vec{r}_2, t' + t \right),\\
    \tau \left( \vec{r}_1, \vec{r}_2 \right) &\equiv \arg \max \limits_t \left[C \left( \vec{r}_1, \vec{r}_2, t \right) \right],
\end{align}
where $T$ is the duration of the observation. In the time--distance helioseismology, authors often utilised the perturbed travel times $\delta \tau \left( \vec{r}_1, \vec{r}_2 \right),$ defined as the difference between the measured travel time $\tau_{\mathrm{msm}}$ and the modelled travel time $\tau_{\mathrm{model}}$:
\begin{equation}
    \label{eq:dtau_def}
    \delta \tau \left( \vec{r}_1, \vec{r}_2 \right) \equiv \tau_{\mathrm{msm}} \left( \vec{r}_1, \vec{r}_2 \right) - \tau_{\mathrm{model}} \left( \vec{r}_1, \vec{r}_2 \right).
\end{equation}

\subsection{Point-to-annulus averaging geometries}
\label{sect:PtA}

The perturbed travel times are usually measured under the given averaging geometries, which reduce the realisation noise component. In applications widely used in literature, four point-to-annulus (PtA) averaging geometries are distinguished. The temporal cross-correlation is always computed between the observed signal $\psi$ at the given point and the observed signal averaged over the normalised annulus with the radius $\Delta$. The final perturbed travel times are constructed from the combinations of the travel times measured from the central point to the rim of the annulus and the travel times measured in the opposite direction. If the annulus is symmetrical with respect to its centre, we refer to the outflow-inflow (o-i; the difference between $\delta \tau \left( \vec{r} = \vec{r}_1, \Delta = \norm{\vec{r}_1 - \vec{r}_2} \right)$ and $\delta \tau \left(\Delta, \vec{r}\right)$) and the mean (mn; the mean value of the two travel times) averaging geometries. The annulus values can be weighted with the cosine (east-west geometry; e-w) or the sine (north-south geometry; n-s) of the polar angle. These geometries are sensitive to the direction of the wave trajectory. Only the differences between the two travel times are computed for the e-w and the n-s geometries.

\ocfigure{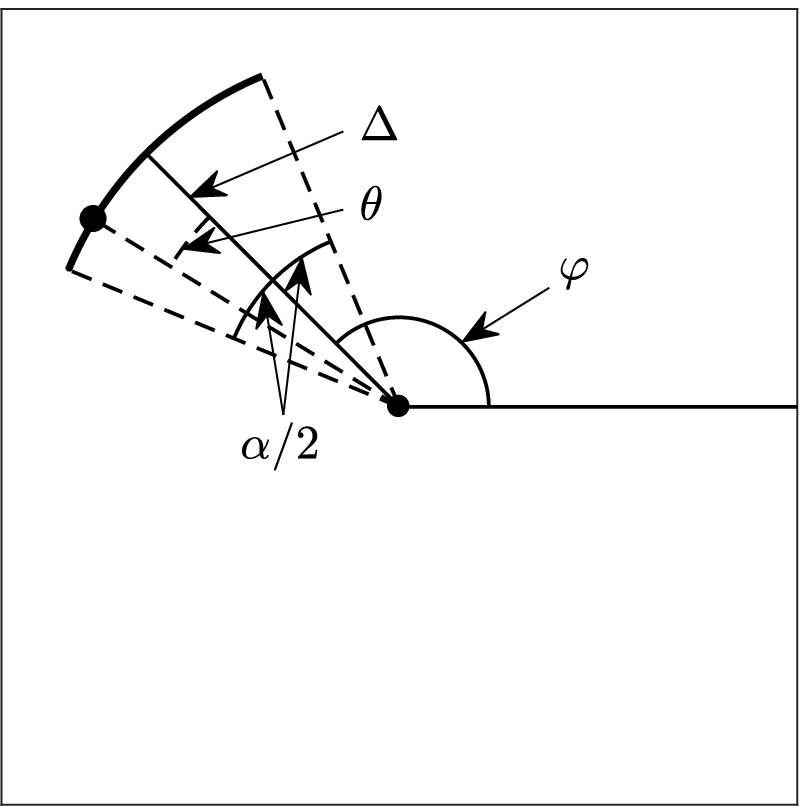}
{Simplified drawing of an arc geometry with the defining parameters $\alpha$ (angular width of the arc), $\varphi$ (inclination of the arc centre from the reference direction), and $\Delta$ (distance between the central point and the arc). The points on the arc are multiplied by a cosine envelope $\cos\left(\theta \pi / \alpha\right)$.}
{fig:definition}

\tcfigure{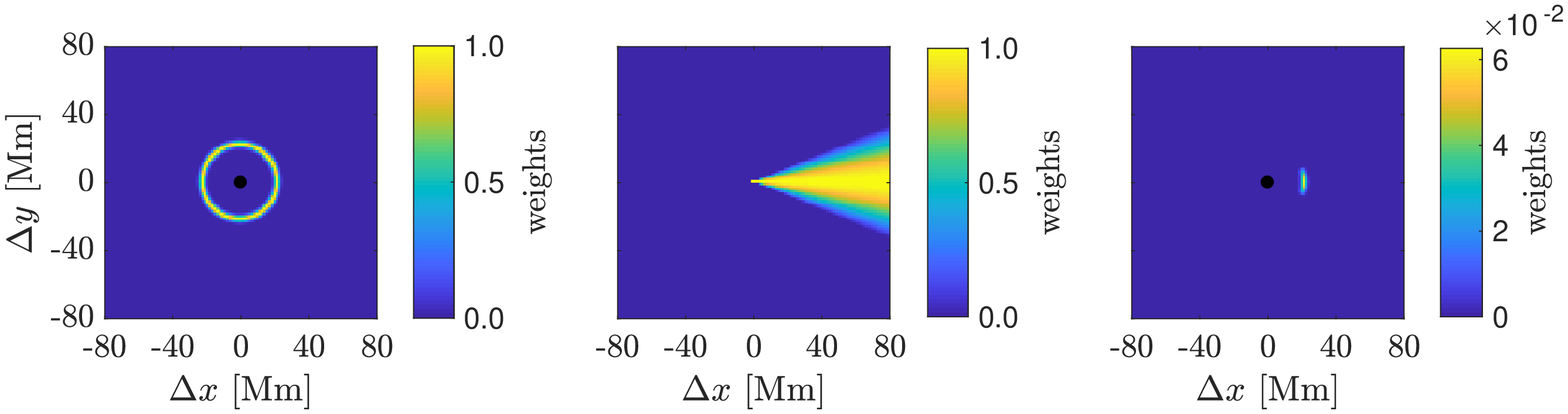}
{Individual steps in the construction of the arc geometry defined by $\alpha = \pi/4$, $\varphi = 0$, and $\Delta \approx 22$~Mm. Left: Unit Gaussian annulus with the radius $\Delta$ and the given \FWHM{}. Middle: Weighted angle interval $\left\langle \varphi - \alpha/2; \varphi + \alpha/2 \right\rangle$. Right: Resulting one-sided arc averaging geometry. The black dots indicate the central points and are not parts of the averaging geometries. For display purposes, the \FWHM{} of the Gaussian annulus was enlarged.}
{fig:arc_construction}

\subsection{Point-to-arc averaging geometries}

The PtA averaging geometries are very useful in quiet-Sun regions because the statistical properties of waves with the same phase speed and generated at the same point are comparable regardless of the direction of propagation. If this is not true, the PtA averaging geometries average regions that are not similar. Hence, the PtA geometries do not fit the real physical conditions in such regions well.

For this reason, we introduce new averaging geometries: one-sided point-to-arc (PtArc) averaging geometries. They represent a natural transition between noisy one-sided point-to-point travel-time geometries and omnidirectional PtA travel-time geometries with a reasonable signal-to-noise ratio. These geometries do not average the observed signal along all the angles but only over small arcs. Therefore, the travel time measured at the edge of a region with different properties is correctly split into travel times inside and outside this region (see Fig.~\ref{fig:ave_geoms}). With the PtArc averaging geometries we can, in principle, invert for plasma properties around active regions as a function of depth, which is highly important. Theoretically, the ideal solution is to use very narrow arcs, or, equivalently, point-to-point travel times defined by Eq.~(\ref{eq:dtau_def}); however, the user must deal with increasing contribution from the realisation noise. Additionally, many independent observations are needed to obtain the same quantity of information, and the inversion process is slowed down.

\subsubsection*{Construction}

One particular arc geometry is defined by three numbers; the angular width of the arc $\alpha$, the orientation of the arc axis $\varphi$, and the distance $\Delta$ (simplified draw is plotted in Fig.~\ref{fig:definition}). The geometry is constructed in the following way:
\begin{enumerate}
    \item select the three numbers,
    \item construct the unit Gaussian annulus with the given \FWHM{} at the distance $\Delta$ from the central point,
    \item construct a function that is equal to one inside the angle interval $\left\langle \varphi - \alpha/2; \varphi + \alpha/2 \right\rangle$ and zero elsewhere,
    \item weight the function with $\cos\left(\theta \pi/\alpha\right)$, where $\theta$ is the inclination of a point of the function with respect to the reference direction defined by $\varphi$,
    \item multiply the Gaussian annulus and the weighted function, and then
    \item the arc geometry is attained via normalisation of the previous multiplication to unity.
\end{enumerate}
For the given $\alpha$ and $\Delta,$ we constructed a set of arcs directing to a set of polar angles $\varphi$, usually with a critical sampling given the $\alpha$. The construction of the arc geometry defined by $\alpha = \pi/4$, $\varphi = 0$, and $\Delta \approx 22$~Mm is shown in Fig.~\ref{fig:arc_construction}. We note that the annulus geometries can be reconstructed from the arc geometries only if $\alpha = \pi$; for instance, the annulus geometry plotted in Fig.~\ref{fig:ave_geoms} corresponds to the combination of two arc geometries defined by $\alpha_1 = \pi$, $\varphi_1 = 0$, $\Delta_1 \approx 29$~Mm and $\alpha_2 = \pi$, $\varphi_2 = \pi$, $\Delta_2 \approx 29$~Mm.

We tested the methodology with the following set of numbers: $\alpha = \pi/4$, $\varphi$ from 0 to $2\pi-\alpha$ with the step $\alpha$ (eight geometries overall), the \FWHM{} of the Gaussian annulus of approximately 1.46~Mm, and $\Delta$ from about 7.3~Mm to about 29.2~Mm with the step of about 1.46~Mm. The selected angular width and orientations covered all directions. The step $\alpha$ corresponded to the smallest value which enabled to enforce the given symmetries of a reference model without an interpolation.


\section{Forward and inverse models}

Using an approximate solution of the wave equation in a perturbed environment we are able to compute the sensitivity kernels $K$. The sensitivity kernels quantify the changes in travel times under the given changes in the reference model. In general, the sensitivity kernels differ by reference model and with a level of approximation. We utilised the model~S \citep{Model_S} as the reference model and the Born approximation \citep{GB02}. The flow and sound-speed sensitivity kernels were computed according to \citet{BG07} and \citet{B04}.

The perturbations of travel times $\delta \tau$ can be computed as
\begin{equation}
    \delta \tau^a \left(\vec{r}\right) = \int \limits_{\sun} \ddif{r'} \dif{z} \sum \limits_{\beta = 1}^P K^a_{\beta}\left(\vec{r'} - \vec{r}, z\right) \dq\left(\vec{r'}, z\right) + n^a\left(\vec{r}\right),
    \label{eq:dtau}
\end{equation}
where $\vec{r}$ and $\vec{r'}$ are the horizontal positions, $z$ the vertical position, the upper index $a$ denotes the independent measurements, $P$ is the number of perturbers, and $n^a$ the realisation noise.


The inverse methods deconvolve Eq.~(\ref{eq:dtau}) and teach us about the perturbers $\dq$. Due to the noise term, the deconvolution leads to the approximate solution $\dqinv$. The multi-channel optimally localised averaging method \citep[MC-SOLA;][]{Jackiewicz_2012} applies an approximate inverse method and searches for the approximate solution in the form
\begin{equation}
    \dqinv \left( \vec{r}_0; z_0\right) = \sum \limits_{i = 1}^{N} \sum \limits_{a = 1}^{M} \w_{a} \left(\vec{r}_i - \vec{r}_0;z_0\right)\delta \tau^a\left(\vec{r}_i\right),
\end{equation}
where $\vec{r}_i$ and $\vec{r}_0$ are the horizontal positions, $z_0$ the target depth, $N$ the number of horizontal positions,  and $\w_{a}$ the unknown weight functions (inverse filters) that must be determined. In our inverse models, the weight functions minimise the $\chi^2$ functional in the form:
\begin{align}
    \chi^2 &= \int \limits_{\sun} \ddif{r'} \dif{z} \sum\limits_\beta \left[\akern \left(\vec{r'}, z; z_0\right) - \mathcal{T}^{\alpha}_{\beta}\left(\vec{r'}, z; z_0\right)\right]^2 + \nonumber\\
    &+ \mu \sum \limits_{i,\,j,\,a,\,b} \w_a \left(\vec{r}_i; z_0\right) \lab \left(\vec{r}_i - \vec{r}_j\right)\w_b \left(\vec{r}_j; z_0\right) + \nonumber\\
    &+ \nu \sum \limits_{\beta \neq \alpha} \int \limits_{\sun} \ddif{r'} \dif{z} \left[\akern \left(\vec{r'}, z; z_0\right)\right]^2 +\epsilon \sum \limits_{a,\,i} \left[\w_a \left(\vec{r}_i; z_0\right)\right]^2 + \nonumber \\
    &+ \sum \limits_{\beta} \lambda^{\beta} \left[\int \limits_{\sun} \ddif{r'} \dif{z} \akern \left(\vec{r'}, z; z_0\right) - \delta^{\alpha}_{\beta}\right],
    \label{eq:chiSOLA}
\end{align}
where $\mu$, $\nu$ and $\epsilon$ are the trade-off parameters, $\akern$ the averaging kernel, $\mathcal{T}^{\alpha}_{\beta}$ the target function, $\lab$ the noise covariance matrices, $\lambda^{\beta}$ the Lagrange multipliers, and $\delta^{\alpha}_{\beta}$ the Kronecker delta function. We invite the reader to consult \citet{Korda_2019a} for more details about the implementation.


\tcfigure{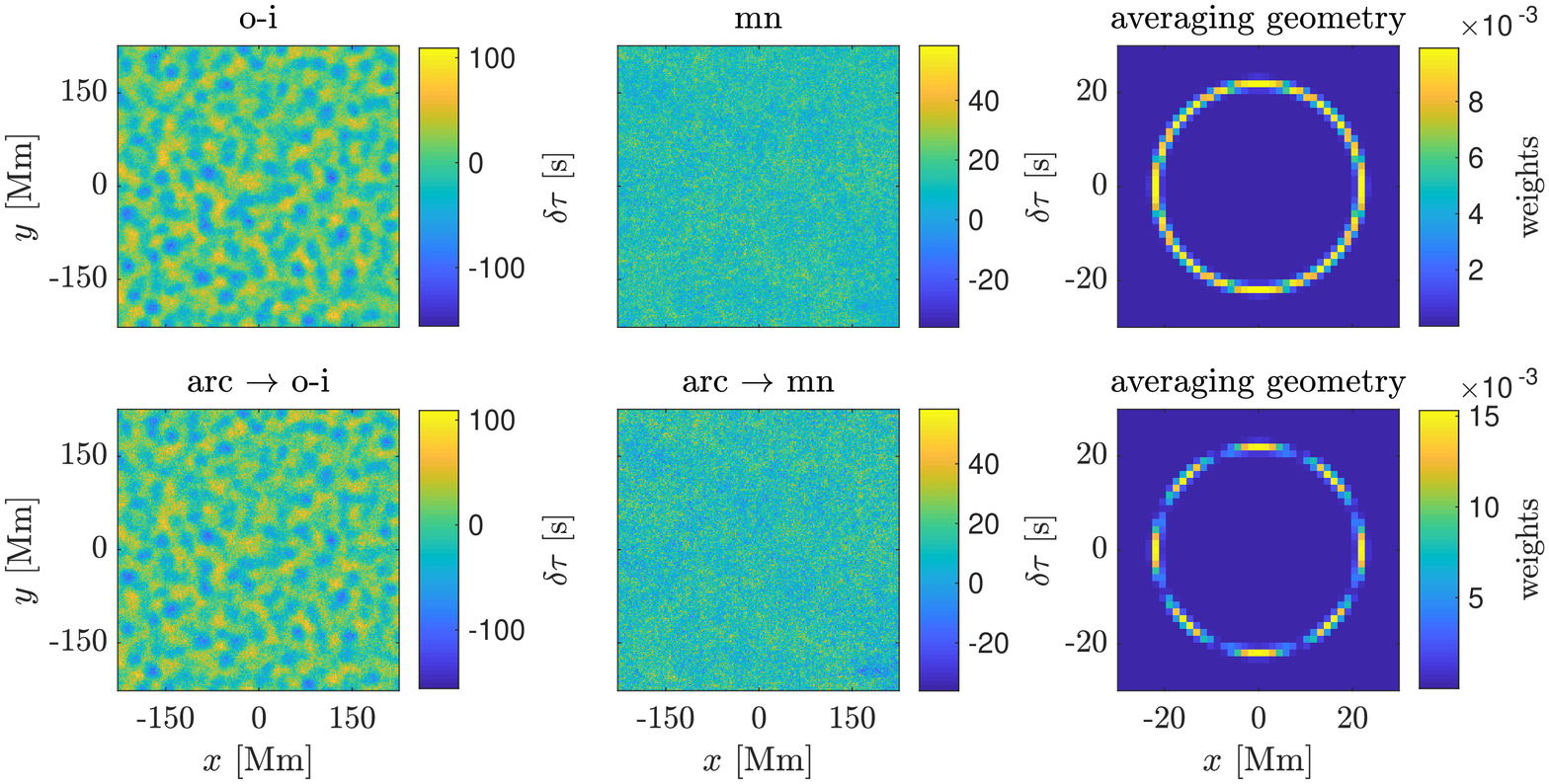}
{Annulus outflow-inflow and mean $f$-mode travel times with $\Delta \approx 22$~Mm. Top row: Annulus geometries; GB04 approach. Bottom row: Reconstruction with the arc geometries; GB02 approach.}
{fig:arc_to_oi}

\section{Consistency check}

The validity regions for both PtA and PtArc travel-time measurements overlap in quiet-Sun regions. We utilised the aforementioned inverse method and perform consistency checks. We tested the consistency of the new one-sided arc geometries in two ways. First, we approximated the PtA travel times from the specific combination of the PtArc travel times. This reconstruction was a consistency check of the new PtArc travel times. Second, we compared the inversions for horizontal flows in a quiet-Sun region based on PtA and PtArc travel times. In quiet-Sun regions, the results should be very similar, hence, this was a check of the whole PtArc pipeline. We note that if the travel times were consistent but the pipeline was not, the issue must have been in the sensitivity kernels.


\subsection{Travel times}
\label{sec:tt_consistency}

Due to our selected $\alpha \neq \pi$, the PtA travel times cannot be fully reconstructed from the PtArc travel times. For the reconstruction, we utilised the Dopplergrams measured with SDO/HMI \citep{HMI1, HMI2} on 8~January 2011. The Dopplergrams were averaged over 24~hours, and the central part of the observed map was at the Carrington coordinates of $156.6^{\circ}$ longitude and $0.0^{\circ}$ latitude (the disc centre was at the coordinates of $156.6^{\circ}$ longitude and $-3.8^{\circ}$ latitude). From the Dopplergrams, we measured the travel times in the PtA and the PtArc averaging geometries. In order to compare the travel times, we combined the PtArc travel times in the following way:
\begin{enumerate}
    \item measure the PtArc travel times from the central point to the rim of the arc and in the opposite direction,
    \item for the PtArc travel times with the given $\varphi$, multiply the travel times with the following: one in the case of o-i and mn geometries; $\cos\left(\varphi\right)$ in the case of the e-w geometry; $\sin\left(\varphi\right)$ in the case of the n-s geometry, and sum up the results,
    \item normalise the result so that the corresponding averaging geometries have the same integrals for positive and negative parts,
    \item subtract the `opposite-direction' travel times (o-i, e-w, n-s) and compute the mean of the `opposite-direction' travel times (mn).
\end{enumerate}

In Figs.~\ref{fig:arc_to_oi} and \ref{fig:arc_to_ewns}, we show the exact GB04 annulus travel times (top rows) and the reconstructed annulus travel times from the GB02 one-sided arc travel times (bottom rows). We note that we computed the $f$-mode travel times with $\Delta \approx 22$~Mm. The corresponding sub-plots have the same colour bars. In the right panels, there are the corresponding averaging geometries. As one can see, the reconstruction could not be perfect because the averaging geometries were slightly different. Because of the cosine and the sine weighting, only six arc geometries were used in the cases of the e-w and the n-s travel times, while all eight in the cases of the o-i and the mn travel times. The statistical properties of the comparison are summarised in Table~\ref{tab:arc_to_annulus}, namely the correlation coefficients between the corresponding travel times and the root mean square ($\rms$) of the differences between the corresponding travel times. 

Histograms of the arc-to-annulus and the corresponding annulus travel times are plotted in Figs.~\ref{fig:arc_hist} and \ref{fig:arc_hist2}. The bin widths are 5~seconds in all histograms. In the histograms in Fig.~\ref{fig:arc_hist}, the black bars indicate the arc-to-annulus travel times, and the yellow bars represent the annulus travel times. The larger widths of the histograms of the arc-to-annulus travel times are probably caused by lower averaging. The black lines in Fig.~\ref{fig:arc_hist2} have unit slopes. The slopes of the linear regressions are 0.89 (o-i), 0.82 (mn), 0.89 (e-w), and 0.90 (n-s). Except the histograms of the mean travel times, the other histograms are centred close to $\delta \tau = 0$~s. The histogram of the mean travel times is centred at about $\delta \tau = 12$~s. This shift is most probably due to the difference of the plasma properties in the near-surface layers of the Sun (roughly the last 1~Mm, where the $f$-mode is significant) as compared to the model S, which is a reference for our travel-time measurement. This systematic offset does not affect the difference travel times, because it is automatically subtracted. It is important to note that the offset in the mean travel times is the same for both the PtA travel times and the equivalent travel times derived from our PtArc measurements.

\tcfigure{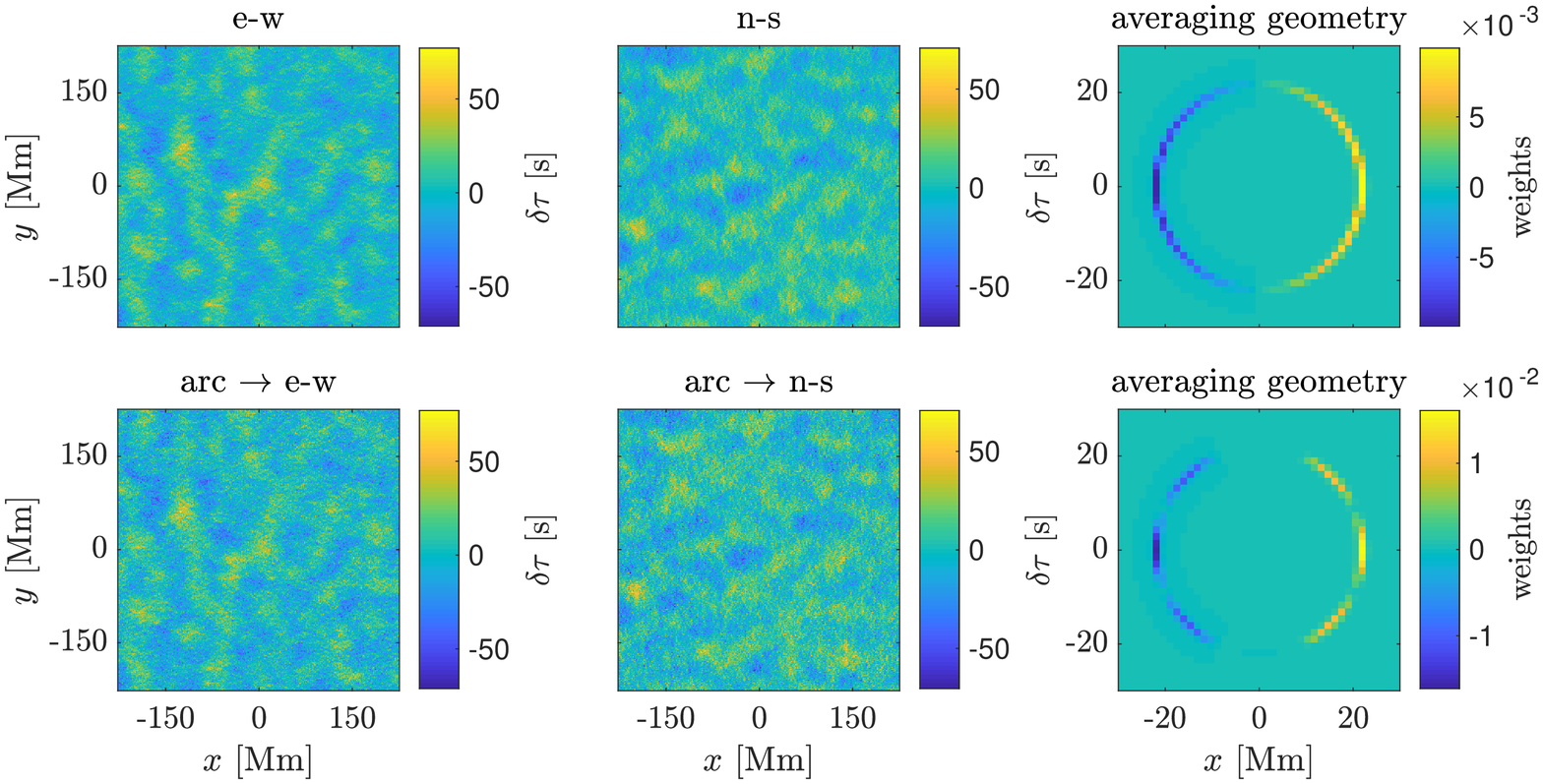}
{Annulus east-west and north-south $f$-mode travel times with $\Delta \approx 22$~Mm. Top row: Annulus geometries; GB04 approach. Bottom row: Reconstruction with the arc geometries; GB02 approach. Only east-west exact and reconstructed averaging geometries are shown.}
{fig:arc_to_ewns}

\begin{table}
    \caption{Correlation coefficients between the corresponding travel times and $\rms$ of their difference. Top part: Computed from the measured travel times (different averaging geometries imply different realisation noises). Bottom part: Travel times were smoothed with the Gaussian (standard deviation of about 4~Mm).}
    \label{tab:arc_to_annulus}
    \centering
    \begin{tabular}{l r r}
        \hline\hline
        \multicolumn{1}{c}{type} & 
        \multicolumn{1}{c}{$\corr$} & 
        \multicolumn{1}{c}{$\rms$} \\
        \multicolumn{1}{c}{} & 
        \multicolumn{1}{c}{} & 
        \multicolumn{1}{c}{[s]} \\
        \hline
        \multicolumn{3}{c}{different noise components} \\
        \hline
        (arc $\rightarrow$ o-i), o-i & 0.89 & 14.4 \\
        (arc $\rightarrow$ mn), mn & 0.68 & 7.4 \\
        (arc $\rightarrow$ e-w), e-w & 0.77 & 10.8 \\
        (arc $\rightarrow$ n-s), n-s & 0.76 & 11.0 \\
        \hline
        \multicolumn{3}{c}{noise components suppressed} \\
        \hline
        (arc $\rightarrow$ o-i), o-i & 0.99 & 2.9 \\
        (arc $\rightarrow$ mn), mn & 0.81 & 1.6 \\
        (arc $\rightarrow$ e-w), e-w & 0.97 & 2.2 \\
        (arc $\rightarrow$ n-s), n-s & 0.96 & 2.5 \\
        \hline
    \end{tabular}
\end{table}

\ocfigure{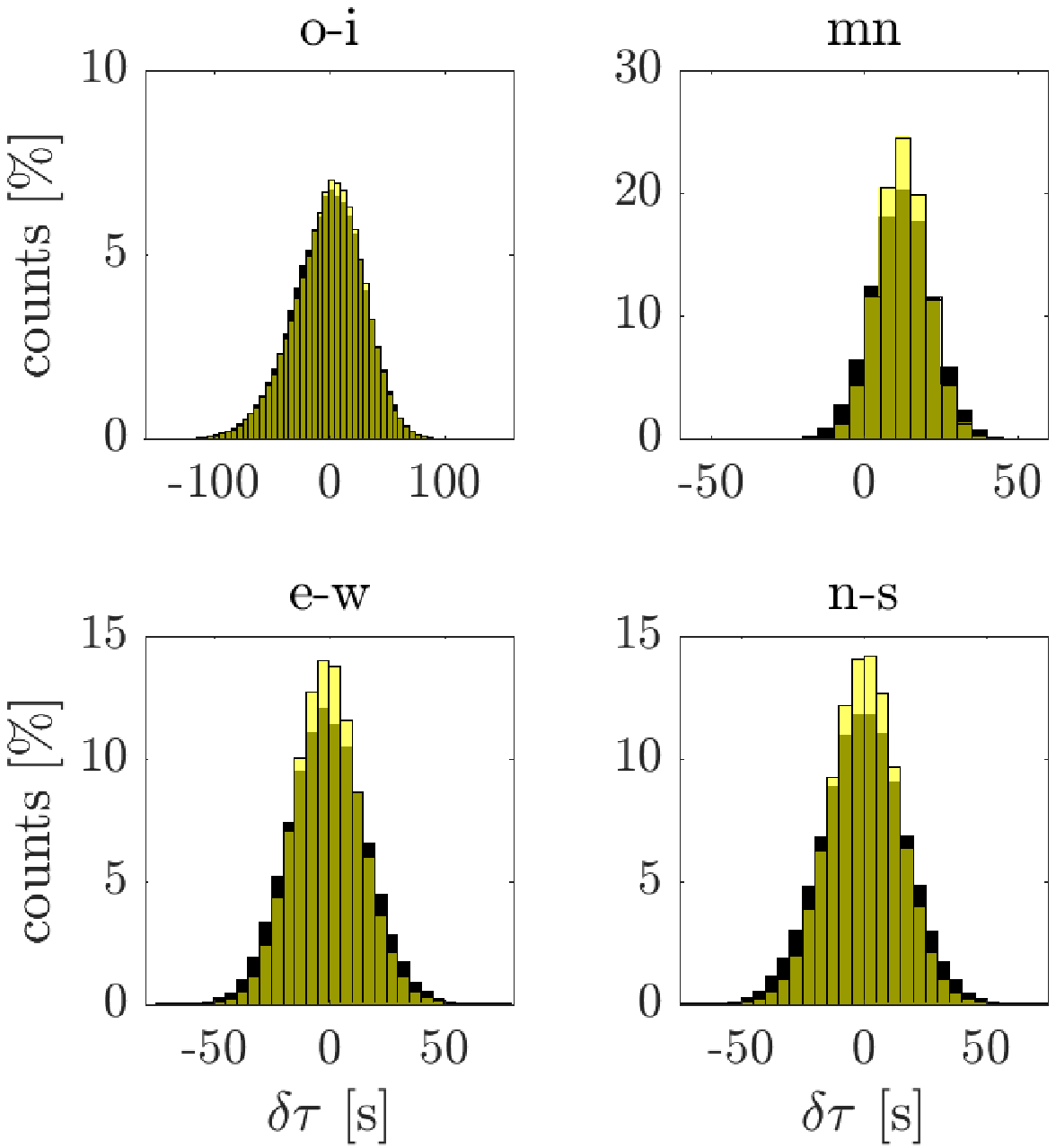}
{Histograms of the corresponding annulus and arc-to-annulus travel times. The black bars indicate the arc-to-annulus travel times. The yellow bars indicate the annulus travel times. The bin widths are 5~s. The total number of counts is 96100.}
{fig:arc_hist}

\ocfigure{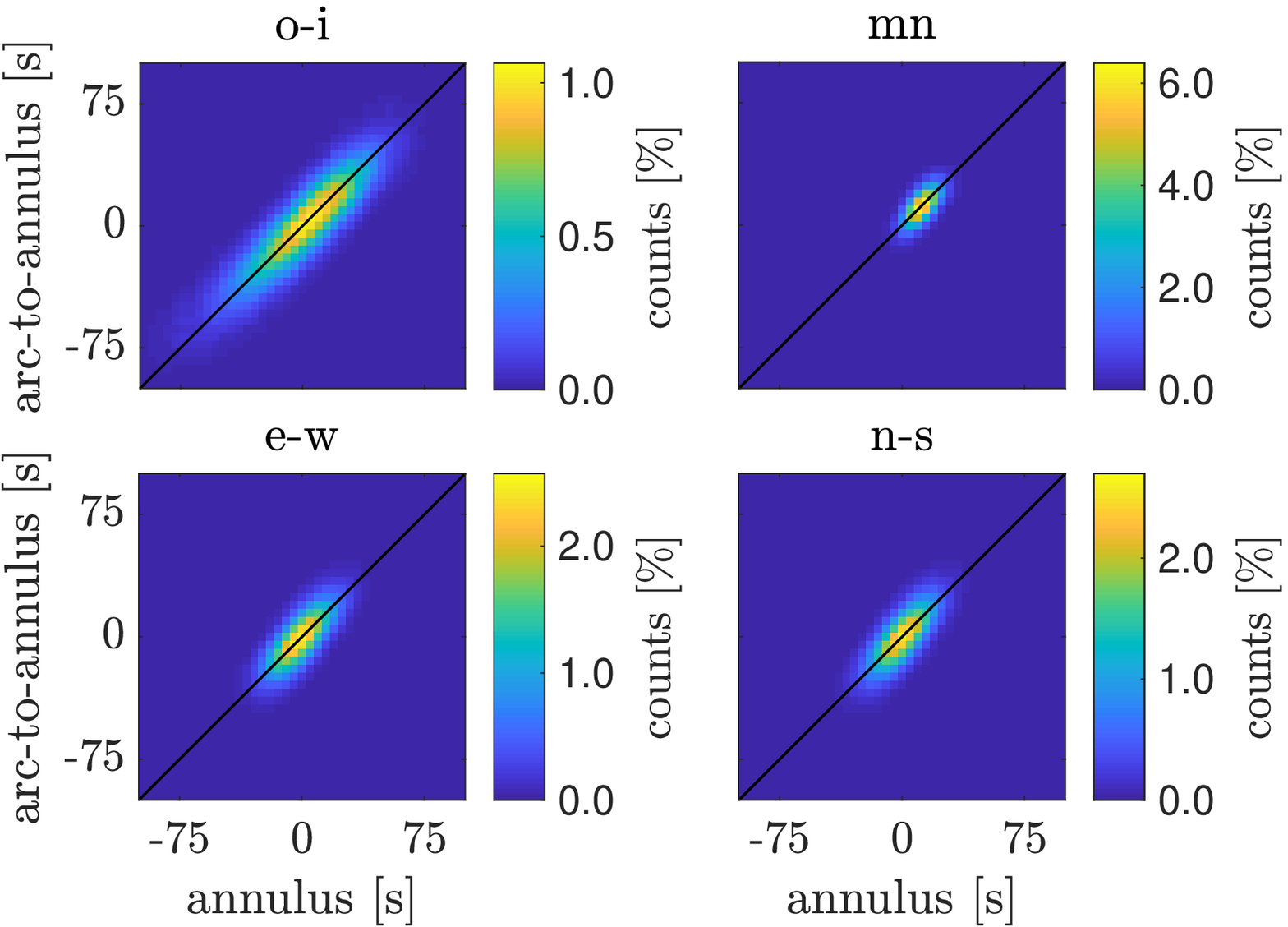}
{2-D histograms of the corresponding annulus and arc-to-annulus travel times. The bin widths are 5~s. The slopes of the black lines are one. The slopes of linear regressions are about 0.9.}
{fig:arc_hist2}

In the cases of the o-i and the mn geometries, the reconstructed travel times are very similar to the exact annulus travel times. Not only are the structures of the travel times nearly the same, but the amplitudes are also comparable. The correlation coefficient of the two averaging geometries (right panels in Fig.~\ref{fig:arc_to_oi}) is 0.90; therefore, the correlation coefficients between the corresponding travel times should not be larger than this value because of the different realisation noise contributions. The correlation coefficients and the $\rms$ values of the differences between the o-i travel times are 0.89 and 14.4~seconds, respectively. In the case of the mean travel times, the numbers are 0.68 and 7.4~seconds. A slightly lower correlation is due to a centre-to-limb trend in the mean travel times. This trend is caused by foreshortening, and the technique of its subtraction for the arc travel times \citep[which is different from the technique for the annulus travel times; see Appendix A of][]{Korda_2021a} can be found in Appendix~\ref{app:trend}. Most of the differences in the corresponding travel times are caused by the different realisation noise components. We smeared the travel times with the Gaussian with the standard deviation $\sigma$ of about 4~Mm. Then the correlation coefficients and the $\rms$ values are 0.99 and 2.9~seconds (o-i) and 0.81 (0.96 if we ignore the trend) and 1.6~seconds (mn).

The e-w and the n-s averaging geometries and travel times are plotted in Fig.~\ref{fig:arc_to_ewns}. The differences between the corresponding travel times are slightly larger because of the larger differences between the corresponding averaging geometries. However, the correlation coefficients are 0.77 (e-w) and 0.76 (n-s). The $\rms$ of the differences between the corresponding travel times are about 11~seconds. After we smeared the travel times with the Gaussian, the correlation coefficients are 0.97 (e-w) and 0.96 (n-s), and the $\rms$ are 2.2~seconds (e-w) and 2.5~seconds (n-s).


\subsection{Horizontal flows}

In the past, our `PtA' horizontal flows in quiet-Sun regions were compared with various types of independent methods \citep[see e.g.][]{SvandaRoudier_2013, Korda_2019b}. Therefore, we can validate the new arc geometries via a comparison of the inverted horizontal flows in quiet-Sun regions with our PtA inversions.

We performed two sets of surface ($f$-mode) inversions for horizontal flows. The first set was based on the PtA averaging geometries, while the other one was based on the new PtArc averaging geometries. Similarly to the approach taken in the previous section, we utilised the Dopplergrams measured at the quiet disc centre on 8~January 2011. The inverted horizontal flows are plotted in Fig.~\ref{fig:inversion_check}. In the top row, we show the inversions for longitudinal flows, and in the bottom row we show the inversions for the latitudinal flows. In the left column, one can see the `GB04, annulus' horizontal flows, while in the right column the new one-sided `GB02, arc' horizontal flows are documented. Even though a different data averaging and slightly different averaging kernels and realisation noises were used, the results in both rows are nearly the same. The larger amplitude of the PtArc horizontal flows corresponds to the larger amount of longer PtArc travel times (see Fig.~\ref{fig:arc_hist}) and is likely caused by lower averaging of the input data and differences between the averaging kernels. In Table~\ref{tab:inversion_check}, one can find correlation coefficients, $\rms$ and $\mean$ of the `arc' minus `annulus', and the slope $s$ of a linear fit with an arc model as an independent variable (so $\mathrm{PtArc} \times s \approx \mathrm{PtA}$).

\ocfigure{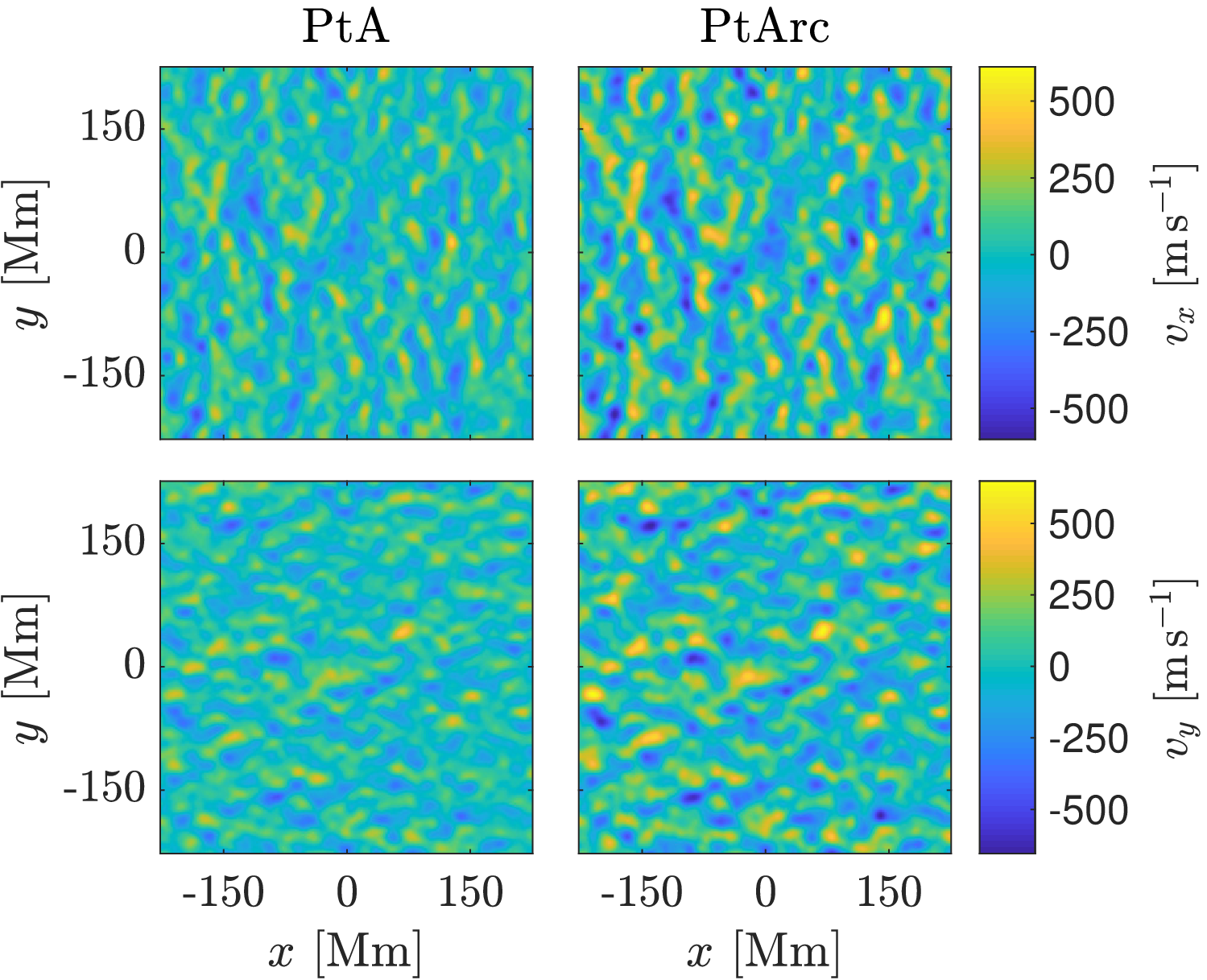}
{Inversions for horizontal flows in the quiet-Sun region at the disc centre on 8~January 2011. Top row: Longitudinal flows. Bottom row: Latitudinal flows. Left column: Inversions in the PtA geometries. Right column: Inversions in the PtArc geometries.}
{fig:inversion_check}

\begin{table}
    \caption{Correlation coefficients between the corresponding inversions for horizontal flows, $\rms$ and $\mean$ of the `arc' minus `annulus', and the slope between them.}
    \label{tab:inversion_check}
    \centering
    \begin{tabular}{c c c c c}
        \hline\hline
        \multicolumn{1}{c}{} & 
        \multicolumn{1}{c}{$\corr$} & 
        \multicolumn{1}{c}{$\rms$} &
        \multicolumn{1}{c}{$\mean$} &
        \multicolumn{1}{c}{$s$} \\
        \multicolumn{1}{c}{} & 
        \multicolumn{1}{c}{} & 
        \multicolumn{1}{c}{[\mps{}]} &
        \multicolumn{1}{c}{[\mps{}]} &
        \multicolumn{1}{c}{} \\
        \hline
        $v_x^{{\rm inv}}$ & 0.96 & 68 & 5 & 0.68\\
        $v_y^{{\rm inv}}$ & 0.95 & 71 & 1 & 0.65\\
        \hline
    \end{tabular}
\end{table}

\ocfigure{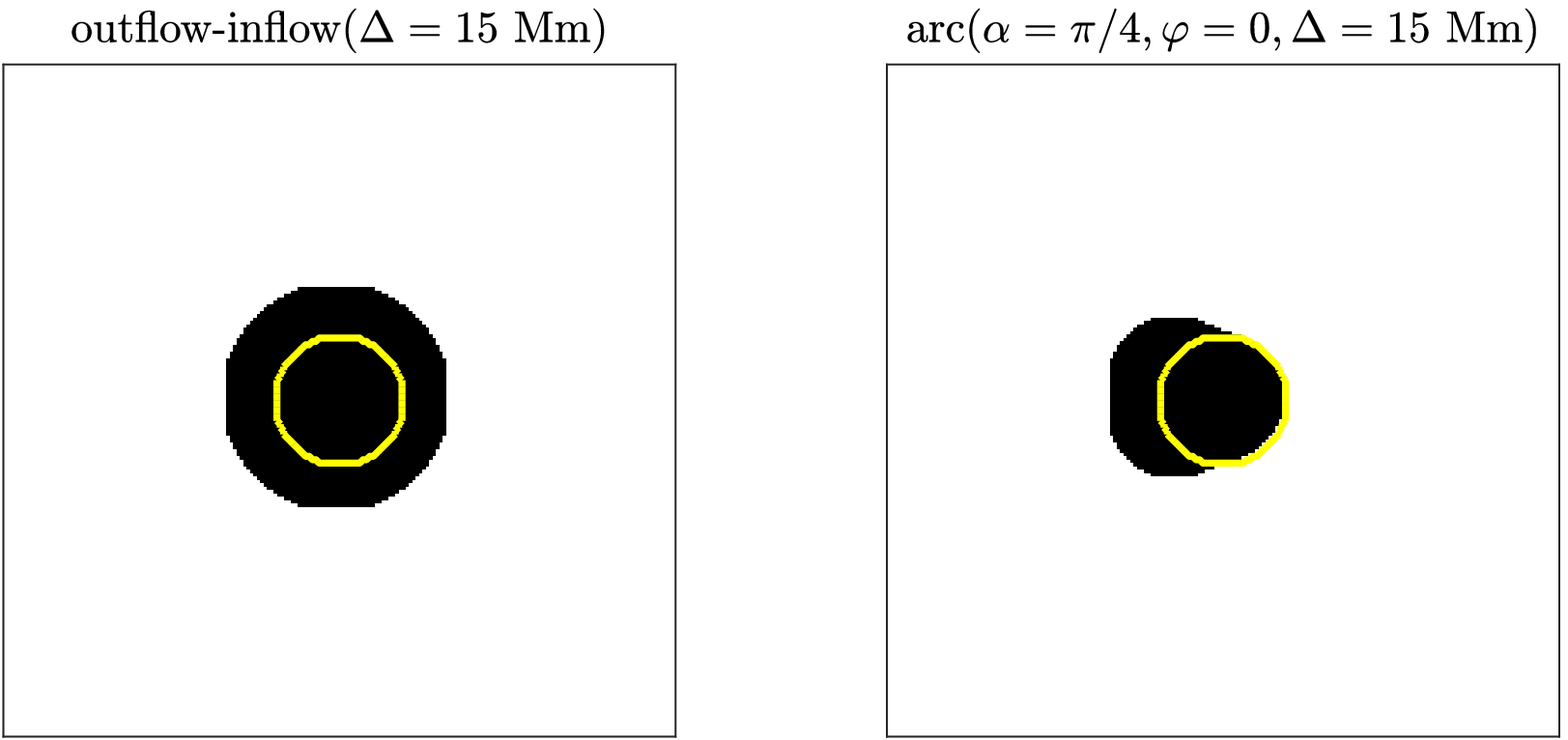}
{Regions containing points that averaged the quiet-Sun and active regions. The yellow contour represents the edge of the active region. Left: Outflow-inflow annulus geometry. Right: Arc geometry defined with $\alpha = \pi/4$ and $\varphi = 0$.}
{fig:geoms_damage}


\section{Test on a circular active region}

In the previous section, we show that the new arc travel-time geometries provide us with very similar results to the annulus geometries in the quiet-Sun regions. The advantage of the arc geometries emerges close to active regions.

For simplicity, we assume a circular sunspot with the radius $\rho$ and the centre at the position $\vec{r}_{{\rm s}}$, the averaging geometry with the distance $\Delta$, and the position of the measurements $\vec{r}_{{\rm m}}$. Then, in the case of the annulus averaging geometries, the affected measurements are inside the disc defined by $\norm{\vec{r}_{{\rm s}} - \vec{r}_{{\rm m}}} \leq \rho + \Delta$. In the case of the one-sided arc geometries, the affected points are inside the ellipse-like figure, whose `major axis' is in the $\varphi$ direction. The lengths of the `semi-major axes' differ in the direction of $\varphi$ and $\varphi + \pi$. In the $\varphi+\pi$ direction, its length is $\rho + \Delta$, similarly to the annulus geometries, but in the direction $\varphi$, the length is $\rho$. The distance $\Delta$ is often greater than 20~Mm; therefore, the combination of the arc geometries with different $\varphi$ parameters should provide us with the unspoiled information about the plasma properties much closer to the sunspot. The damaged parts of the travel-time measurements are visualised in Fig.~\ref{fig:geoms_damage}. In the left panel, we show the affected region for the outflow-inflow annulus averaging geometry, while in the right panel we show the affected region in the case of one of the arc geometries. The yellow central contour corresponds to the edge of an active region. By selecting a set of $\varphi$s, we can `scan' the close vicinity of the spot.

\tcfigure{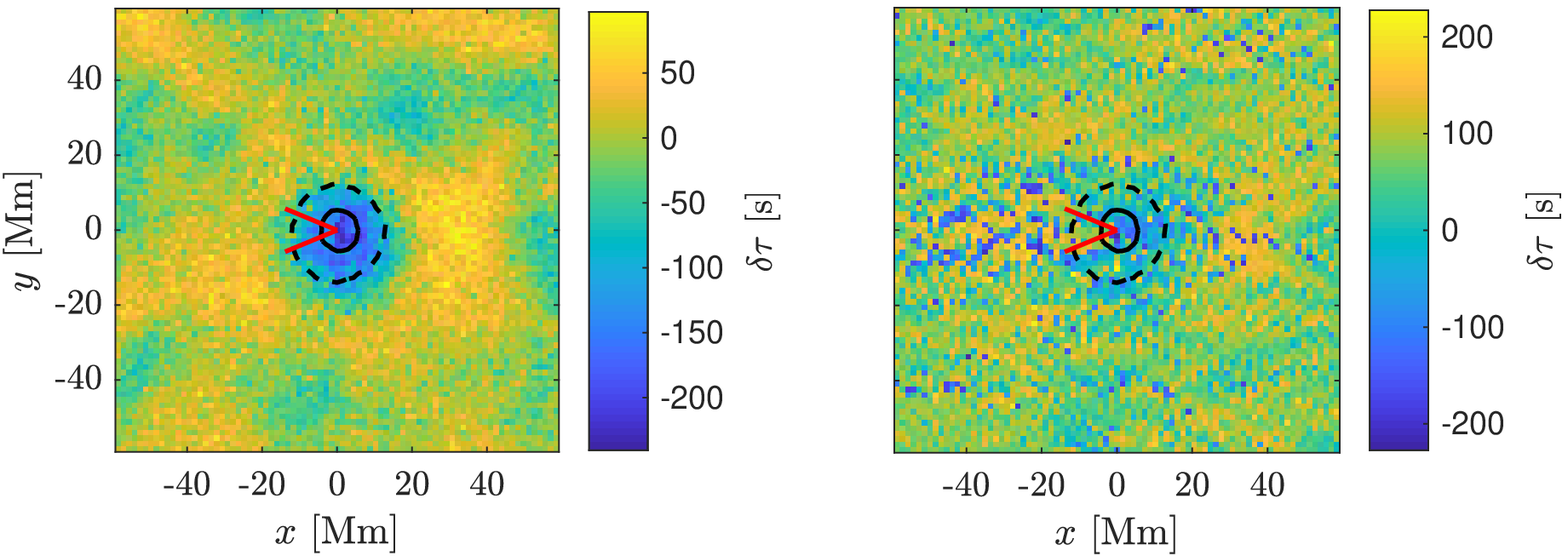}
{GB02 annulus and the arc $f$-mode travel times with $\Delta \approx 29$~Mm measured around the active region NOAA~11084 on 2~July 2010. Left: Outflow-inflow annulus geometry. Right: One-sided arc geometry with $\alpha = \pi/4$ and $\varphi = \pi$. The black contours correspond to umbra (full) and penumbra (dashed). The red wedges show the orientation and the width of the arc geometry.}
{fig:tt_arc_vs_ann}

\subsection{Travel times}

For the demonstration, we selected a roundish H-type active region NOAA~11084 observed with SDO/HMI on 2~July 2010 at Carrington coordinates of $144.5^\circ$~longitude and $-19.1^\circ$~latitude (the Dopplergram centre was at the coordinates $143.0^\circ$~longitude and $-20.0^\circ$~latitude; the disc centre was at $137.7^\circ$~longitude and $3.1^\circ$~latitude). We measured the surface-mode travel times with $\Delta \approx 29$~Mm around this active region in both the annulus and the one-sided arc averaging geometries. Both travel times were measured using the GB02 approach. The measured travel times are plotted in Fig.~\ref{fig:tt_arc_vs_ann}. In the left panel, we show the annulus travel time, and in the right panel the arc travel time ($\alpha = \pi/4$, $\varphi = \pi$) is shown. The full central contour roughly surrounds the umbra and the outer dashed contour surrounds the penumbra. The red wedges show the orientation and the width of the arc geometry. In accordance with the estimation, the arc travel time is spoiled, simply said, in the region outside the red wedge. Inside the wedge in the penumbra, the amplitude of the arc travel time does not seem to be reduced by the presence of the magnetic field. Therefore, the one-sided arc travel times in this specific geometry with combination of the GB02 travel-time approach contain undisturbed information about plasma flows in the eastern direction.


\tcfigure{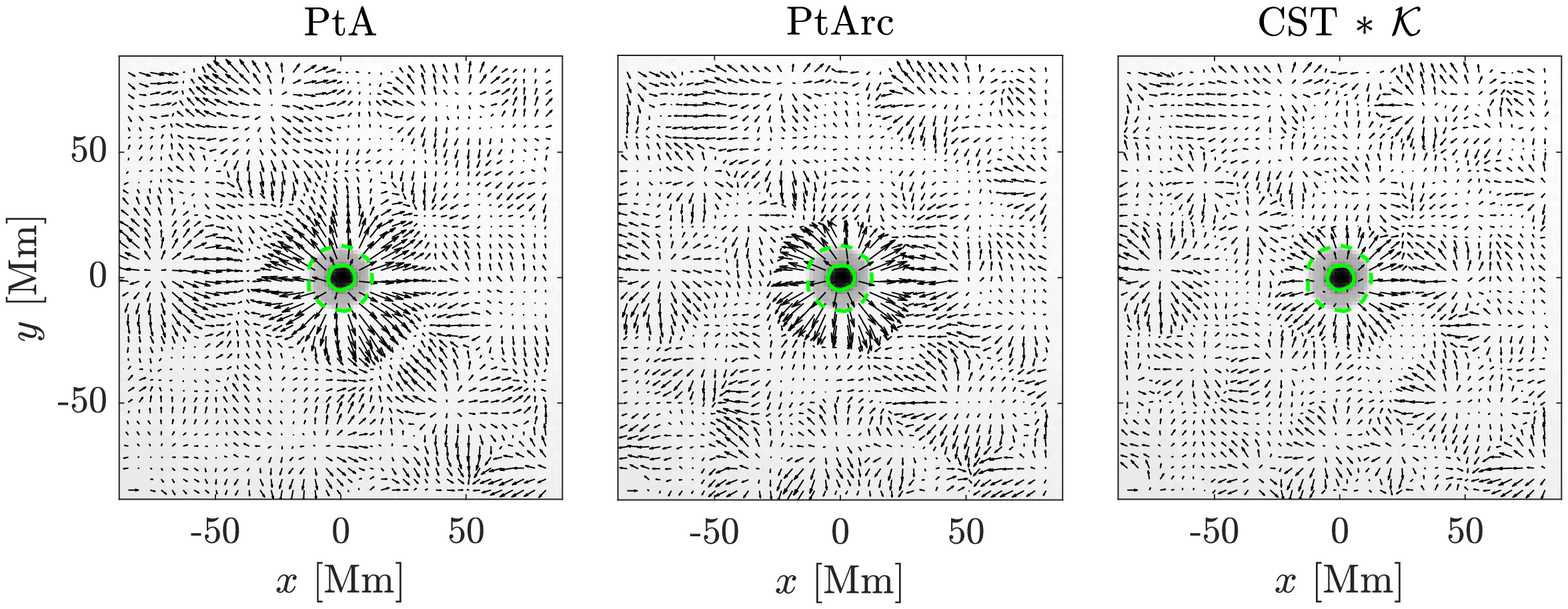}
{Inversions for horizontal flows around the active region NOAA~11084 on 1~July 2010. Left: Inversions in PtA geometries. Middle: Inversions in PtArc geometries. Right: Flows based on CST. The reference arrows at the bottom left corners correspond to 250~\mps{} and all have the same length. The full and the dashed contours correspond to umbra and penumbra. The background images are the associated intensitygrams.}
{fig:inversion_check_spot}

\ocfigure{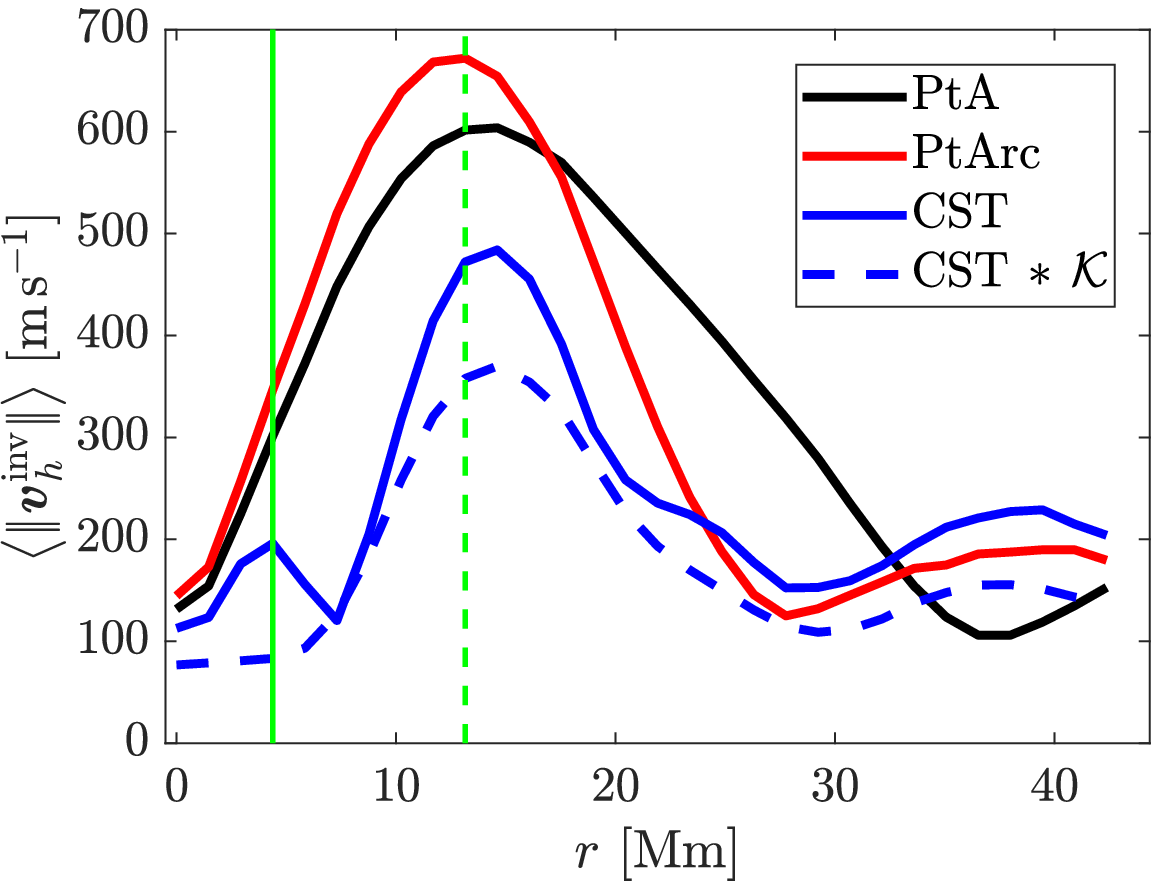}
{Radial profiles of azimuthally averaged speeds. The local minima of the speed are at distances 37.3, 28.1, and 28.5~Mm for PtA, PtArc, and CST respectively. The green vertical lines indicate the edges of umbra and penumbra. The distances correspond to minima of local parabolic fits. We note that to estimate the extent of the moat flow we used raw CST data; otherwise, the edge is systematically shifted by 1~Mm.}
{fig:radial_profile_vh}

\subsection{Horizontal flows}

The surface inversions for the horizontal flows can be directly compared with other methods independent on helioseismology. One of the suitable methods is local correlation tracking, which does not suffer from the magnetic fields. The advanced coherent structure tracking \citep[CST;][]{CST_1, CST_2} is particularly very convenient for the direct comparison of the horizontal flows. These methods are based on tracking granules. Therefore, these methods fail at the edges and within the active regions where the granules are not observed.

To compare the horizontal flows from the CST technique and helioseismic inversions, we need to decrease the spatial resolution of the CST results. The spatial resolution of our helioseismic inversions is quantified by the averaging kernels $\akern$. Therefore, we convolved the CST horizontal flows with corresponding components of the averaging kernels. The CST provides us with surface flows only; therefore, we only utilised the surface cut through the averaging kernel. The convolution with the averaging kernels has one side effect. The CST method utilised the segmentation of granules and their tracking in the co-aligned time-lapse datacubes. In the umbra and penumbra, there are no granules. The structures segmented by the segmentation (such as penumbral grains) have different velocity properties, not to mention that their velocities are only apparent \citep[see e.g.][]{Schlichenmaier_1998}. Hence, the magnitudes of the CST-derived horizontal flows are much smaller (by almost an order of magnitude) within the umbra and penumbra regions and this decrease of the flow magnitude is abrupt on the quiet Sun-penumbra boundary. When smoothing by the averaging kernel, in the moat region the kernel also accumulates the values from the penumbra, which leads to a decrease in the CST-derived flow magnitude. 

The results are plotted in Fig.~\ref{fig:inversion_check_spot}. In each panel, one can see the inversion model under the given approach. The arrows visualise the inverted horizontal flows. The green contours surround the umbra and the penumbra. The background images are the corresponding intensitygrams. The agreement between the middle (PtArc) and the right ($\mathrm{CST} \ast \mathcal{K}$) panels in regions where both methods return valid results provides strong evidence that the new one-sided arc averaging geometries can teach us about the properties of active regions. Using PtArc geometries, we detect a divergent flow whose location and amplitude is comparable with the moat flow. 

In the left panel, we show the horizontal flows derived from the usual PtA inversion applied to the GB02 travel times. A similar approach was also used, for example, by \citet{Svanda_2014}. In the quiet-Sun regions close to the active region, one can see the moat flow with the amplitude of about 500~\mps{}. The measured moat flow is about one-third more extended in comparison to both the PtAr inversion model and the CST model(see Fig.~\ref{fig:radial_profile_vh}). Hence, the divergent moat flow is measured in places where both the PtArc and the CST models changed their signs and no longer detect the moat flow. The larger extension of the moat flow in the PtA flow inversion is likely due to the two-sided travel-time averaging geometries. 

In this regard, we note that \citet{Hughes_2005}, for example, also found that the travel times of the waves are affected by the nearby magnetised region in a region much larger than the size of the active region. It is in agreement with our finding that the PtA inversions are affected in a region larger than the magnetised region itself. The PtArc model, unlike the PtA model, at least provides us with correct information about the extent of the moat flow and perhaps also about the flows up to the edge of penumbra. As a further test, we performed a similar comparison for even simpler flow inversions than PtA presented here in Appendix~\ref{app:inversion}. Those inversions are very straightforward in the interpretation. However, they once again involve the difference (two-sided) travel-time geometries and thus lead to the derived moat-like region, which is more extended than the moat flow measured in PtArc or CST models. We note that in the more distant quiet-Sun regions all the discussed models are comparable.

In both travel-time inversions, the magnitude of the moat flow seems about 50\% larger than that derived by the CST granulation tracking. We note that the surface feature-tracking methods and time--distance inversions do not represent exactly the same depths, as the time--distance inversions aggregate some signal from sub-surface layers. Some numerical simulations \citep[e.g.][]{Rempel_2011} indicate that there might be an increase of magnitude of the flows around the simulated sunspot just under the surface. The decrease of the CST-derived flow magnitude as a side effect of the resolution matching described above also plays a role here.

The statistical comparison between the PtArc and the CST horizontal flows is in Table~\ref{tab:inversion_check_spot}. We note that larger $\rms$ and $\mean$ values were mostly due to the active region and slightly different spatial resolution.

\begin{table}
    \caption{Correlation coefficients between the $\mathrm{CST} \ast \mathcal{K}$ and the `arc' horizontal flows and $\rms$ and $\mean$ of the arc minus $\mathrm{CST} \ast \mathcal{K}$. We note that the parameters were computed from the active region and its quiet surrounding plotted in Fig.~\ref{fig:inversion_check_spot}.}
    \label{tab:inversion_check_spot}
    \centering
    \begin{tabular}{c c c c}
        \hline\hline
        \multicolumn{1}{c}{} & 
        \multicolumn{1}{c}{$\corr$} & 
        \multicolumn{1}{c}{$\rms$} & 
        \multicolumn{1}{c}{$\mean$} \\
        \multicolumn{1}{c}{} & 
        \multicolumn{1}{c}{} & 
        \multicolumn{1}{c}{[\mps{}]} &
        \multicolumn{1}{c}{[\mps{}]} \\
        \hline
        $v_x^{{\rm inv}}$ & 0.86 & 84 & -33 \\
        $v_y^{{\rm inv}}$ & 0.86 & 84 & -34 \\
        \hline
    \end{tabular}
\end{table}

In Tables~\ref{tab:summarise_vx} and \ref{tab:summarise_vy}, we summarise the comparison between the helioseismic inversions for surface horizontal flows and the surface CST measurements. The correlation coefficients between the models are listed above the diagonals. The parameter $s$ corresponds to the slopes of the linear fits of the scatter plots between the given models. The model listed in the row was considered to be the independent variable. Due to slightly different smearing of the models, the fits contained a non-negligible constant parameter. We computed two sets of the $s$ parameters. The first set was computed from the area visualised in Fig.~\ref{fig:inversion_check_spot} after masking out the active region. The slopes are about unity in this case due to most of the area is in quiet Sun. In Tables~\ref{tab:summarise_vx} and \ref{tab:summarise_vy}, these values are in parentheses. The second set was computed from an annulus around the active region. The inner and outer radii of the annulus were about 10~Mm and 37~Mm, respectively. The annulus masked out the imprecise measurements inside the umbra and the inner penumbra and the quiet-Sun region around it. These slopes are outside the parentheses in the two tables.

\begin{table}
    \caption{Summary table of the comparisons between PtA and PtArc helioseismic inversions and the surface $\mathrm{CST} \ast \mathcal{K}$ measurements. Above the diagonal, the correlation coefficients are given. Below the diagonal, the scaling coefficient $s$ corresponds to the slope of a fitted line. The value of $s$ represents the coefficient with which we have to multiply the row in order to obtain the column. The values outside and inside the parentheses were computed from a closer or wider surrounding of the active region, respectively.}
    \label{tab:summarise_vx}
    \centering
    \begin{tabular}{l|ccc}
    \hline\hline
    \backslashbox{$s$}{$\corr$} & $v_{x,\,{\rm PtA}}^{{\rm inv}}$ & $v_{x,\,{\rm PtArc}}^{{\rm inv}}$ & $v_{x,\,{\rm CST}\, \ast\, \mathcal{K}}^{{\rm inv}}$\\
    \hline
    $v_{x,\,{\rm PtA}}^{{\rm inv}}$ &\cellcolor[gray]{0.8} 1.00 & 0.80 (0.79) & 0.78 (0.74) \\
    $v_{x,\,{\rm PtArc}}^{{\rm inv}}$ & 0.92 (0.88) & \cellcolor[gray]{0.8}1.00 & 0.95 (0.87) \\
    $v_{x,\,{\rm CST}\, \ast\, \mathcal{K}}^{{\rm inv}}$ & 1.45 (1.02) & 1.54 (1.06) & \cellcolor[gray]{0.8}1.00 \\
    \hline
    \end{tabular}
\end{table}

\begin{table}
    \caption{Summary table of the comparisons between PtA and PtArc helioseismic inversions and the surface $\mathrm{CST} \ast \mathcal{K}$ measurements. See Table \ref{tab:summarise_vx} for details.}
    \label{tab:summarise_vy}
    \centering
    \begin{tabular}{l|ccc}
    \hline\hline
    \backslashbox{$s$}{$\corr$} & $v_{y,\,{\rm PtA}}^{{\rm inv}}$ & $v_{y,\,{\rm PtArc}}^{{\rm inv}}$ & $v_{y,\,{\rm CST}\, \ast\, \mathcal{K}}^{{\rm inv}}$\\
    \hline
    $v_{y,\,{\rm PtA}}^{{\rm inv}}$ & \cellcolor[gray]{0.8}1.00 & 0.77 (0.76) & 0.66 (0.68) \\
    $v_{y,\,{\rm PtArc}}^{{\rm inv}}$ & 0.87 (0.80) & \cellcolor[gray]{0.8}1.00 & 0.92 (0.88) \\
    $v_{y,\,{\rm CST}\, \ast\, \mathcal{K}}^{{\rm inv}}$ & 1.20 (0.84) & 1.47 (1.03) & \cellcolor[gray]{0.8}1.00 \\
    \hline
    \end{tabular}
\end{table}


\section{Conclusions}

We introduce the one-sided arc averaging geometries in the time--distance local helioseismology. The one-sided geometries allow for inhomogeneities in the observed signal, which can be useful especially in the vicinity of active regions.

The one-sided arc geometries are a superset of the annulus geometries. We selected the specific arc geometries with the width of the arc $\alpha = \pi/4$ and the arc orientations $\varphi$ from 0 to $2\pi-\alpha$ with the step $\alpha$. With these arc geometries, we successfully reconstruct the corresponding annulus travel times. The remaining differences are caused by the different realisation noise components. From the histograms and scatter plots of the corresponding travel times, we note the travel-time histograms are narrower (see Fig.~\ref{fig:arc_hist}) in the case of the annulus geometries. This may be related to the higher level of averaging in the case of the annulus geometries.

We observe the trend in the arc travel times. Similar to the mean travel times, the trend is caused by the foreshortening. The radial centre-to-limb trend in the case of the mean travel times is changed to the band-like trend with the same orientation $\varphi$ as the given arc geometry. We reconstruct the radial trend with the combination of the arc geometries, and we propose and test a trend-subtraction method (see Appendix~\ref{app:trend}).

We compared the horizontal flows inverted under the annulus and the new one-sided arc averaging geometries in the quiet-Sun region, where the inversions of the annulus travel times were tested against various independent measurements. Both inverted results are similar (see Fig.~\ref{fig:inversion_check}); therefore, the new one-sided arc geometries give us the correct results in the quiet-Sun regions.

We demonstrate the advantages of the new one-sided arc geometries in active regions. We tested the inverted horizontal flows around the H-type active region NOAA~11084 using the CST technique. Furthermore, we show that by using this methodology we can successfully measure flows up to the outer penumbra, including the moat flow. Again, both sets of horizontal flows (CST-based and PtArc) are comparable (see Fig.~\ref{fig:inversion_check_spot}). Therefore, we suggest applying the one-sided arc averaging geometries and GB02 travel-time approximation in the vicinity of active regions in order to learn about the depth structure of flows around them. 


\begin{acknowledgements}
D.K. was supported by the Grant Agency of Charles University under grant No. 532217. M.{\v S}. is supported by the project RVO:67985815. M.{\v S}. and D.K. were supported by the grant project 18-06319S awarded by the Czech Science Foundation. This work was granted access to the HPC resources of CALMIP under the allocation 2011-[P1115]. The sensitivity kernels were computed by the code {\sc Kc3} kindly provided by Aaron Birch. This research has made use of NASA Astrophysics Data System Bibliographic Services. The authors would like to thank the anonymous referees for the valuable comments and suggestions, which greatly improved the quality of the paper.
\end{acknowledgements}

\bibliographystyle{aa}
\bibliography{BIBL}


\appendix
\onecolumn

\section{Travel-time trends}
\label{app:trend}

The travel times computed in the arc geometries follow a trend. Similarly to the mean travel times, the trend is caused by geometric distortion due to projecting a sphere onto a plane called foreshortening. Due to narrow arcs, the radial centre-to-limb trend observed in the mean travel times is transformed into a band-like trend with the band orientation $\varphi$ in the cases of the arc travel times. If we transform the travel times measured under the arc geometries into the mean travel time before the subtraction of the trend, we reconstruct the radial centre-to-limb trend (see Fig.~\ref{fig:arc_trend_mn}). The trend has nothing to do with the perturbations of the plasma properties, and hence it must be subtracted. We separated the trend in the following way:

\begin{enumerate}
    \item compute long-term (35~days) average travel times of all the arc geometries,
    \item blur the average travel times with the wide Gaussian (\mbox{$\sigma = 30$~Mm}) to remove small-scale structures and remaining noise,
    \item symmetrise the blurred travel times with respect to the axis of symmetry given by $\varphi$,
    \item fit the parabola to each line parallel to the axis of symmetry and replace the elements of the travel times with the fitted values (do not fit the points that are too close to the edges of the field of view),
    \item blur the constructed image with the narrow Gaussian (\mbox{$\sigma = 3$~Mm}) to remove possible discontinuities,
    \item compute the mean value in the central area and subtract it from the blurred image (foreshortening is zero in the disc centre).
\end{enumerate}
An example of the unspoiled arc travel time and the corresponding trend is plotted in Fig.~\ref{fig:arc_trend}. The band character of the trend is clearly visible in the right panel of the figure.

\tcfigure[!ht]{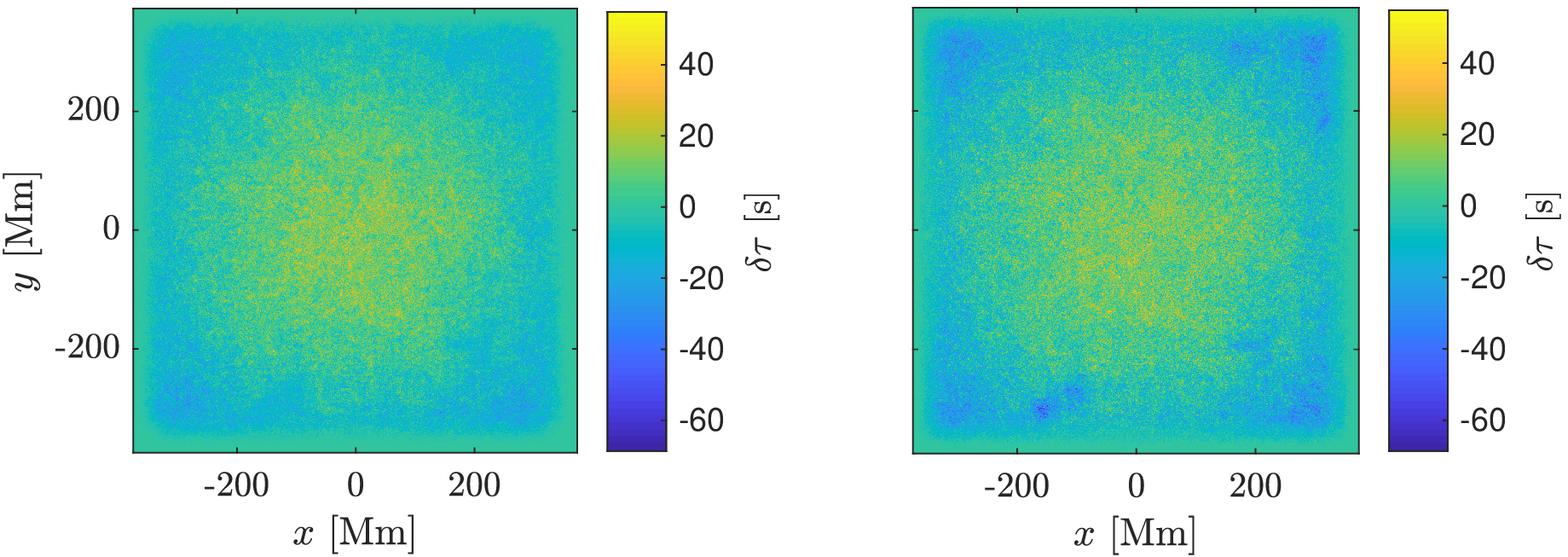}
{Mean travel times with $\Delta \approx 21$~Mm before the subtraction of the trend. Left: Exact GB04 mean travel time. Right: Reconstruction using the GB02 arc travel times. This reconstruction was made of all PtArc travel times (see averaging geometry in Fig.~\ref{fig:arc_to_oi}) and naturally contains a combination of all band-like trends.}
{fig:arc_trend_mn}

\tcfigure[!ht]{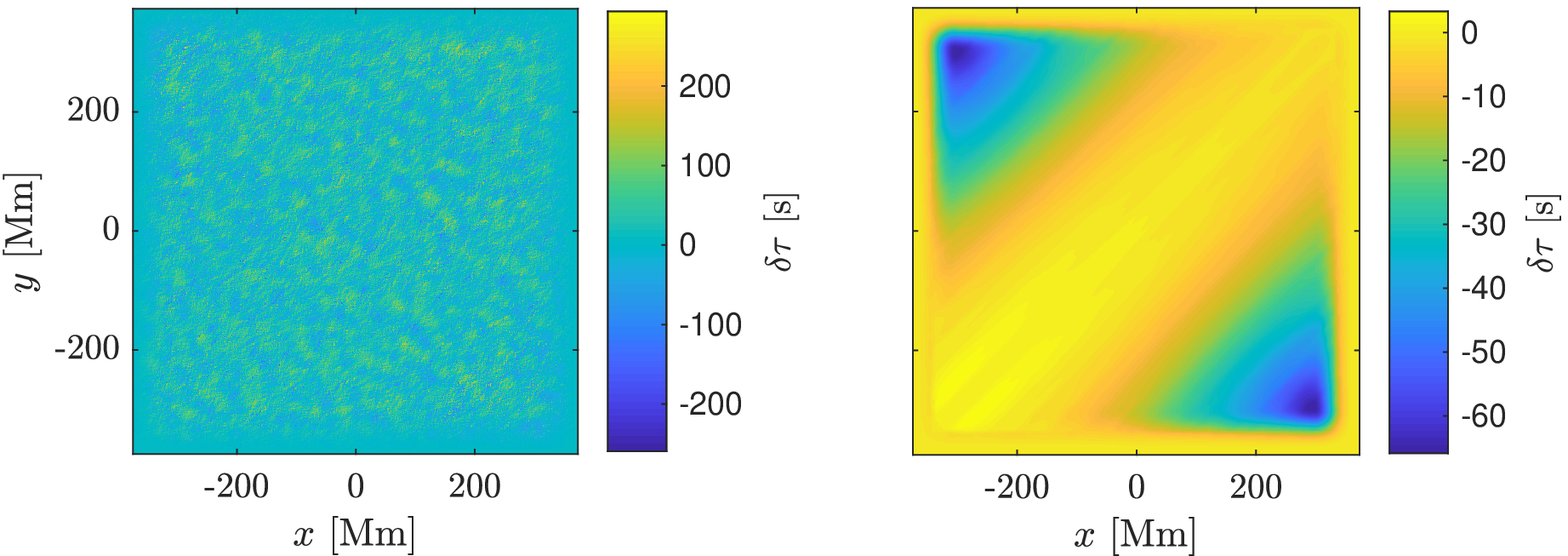}
{GB02 one-sided arc travel time with $\alpha = \pi/4$, $\varphi = \pi/4$ and $\Delta \approx 21$~Mm. Left: Travel time after the trend subtraction. Right: Corresponding trend.}
{fig:arc_trend}

\section{Point-to-quadrant-like inversions}
\label{app:inversion}

To further demonstrate the improvement of our PtArc inversions over the PtA inversions in the magnetised regions, we constructed very simple and very basic point-to-quadrant-like inversions, essentially following the strategy used in the advent of time--distance helioseismology. We limited the set of travel-time geometries to the simplest possible case, essentially following the findings of \citet{Burston_2015}. According to that work, for instance, the east-west component of the horizontal velocity ($v_x$ in our notation) has the strongest impact on e-w travel times (following the notation described in Section \ref{sect:PtA}). Similarly, the north-south component of velocity ($v_y$) has the strongest impact on n-s travel times. These inversions are very simple to interpret. 

Therefore, we constructed inversions involving only those travel-time geometries. Such inversions are, in principle, equivalent to point-to-quadrant flow inversions used, for example, by \citet{Hindman_2004} (their Fig.~6 and related text) and introduced much earlier by \citet{Kosovichev_1997}. Our approach differs from that of \citet{Kosovichev_1997}, \citet{Hindman_2004}, and others, by the fact that our averaging geometry is a smooth function of the azimuthal angle (sine or cosine), whereas they used a step-wise function of the azimuthal angle to separate the east, west, north, and south quadrants. Despite the differences, the principles are similar, and hence we can denote these simplistic inversions as point-to-quadrant (PtQ-1). 

For simplicity and to obtain results directly comparable to the inversions shown in Fig.~\ref{fig:inversion_check_spot}, we once again focused only on the $f$-mode travel times. In fact, our very simplistic PtQ-1 inversions considered a combination of only 16 independent travel-time measurements with varying sizes of the surrounding annulus. This is significantly less than in the case of the PtA inversions presented in Fig.\ref{fig:inversion_check_spot} (48 travel-time measurements combined) and PtArc (128 independent measurements combined). 

The results are shown in Fig.~\ref{fig:inversion_check_spot_appendix}. Due to the lower number of travel-time measurements considered, the signal-to-noise ratio $\mathrm{S/N} \equiv {\sqrt{\rms^2\left(\mathrm{inverted}\right) - \rms^2\left(\mathrm{noise}\right)}}/{\rms\left(\mathrm{noise}\right)}$ is slightly worse in PtQ-1 inversions (the $\mathrm{S/Ns}$ are 6.4 for PtQ-1, 6.6 for PtA, and 20.0 for PtArc inversions, respectively). The averaging kernel (Fig.~\ref{fig:rakern_appendix}) shows that there is a considerable cross-talk from the perpendicular horizontal velocity component (about 25\%), of which the realisation depends on a particular configuration of the flow. For pure divergent flows, these cross-talk contributions act as convergent flows. 

\wfigure[!ht]{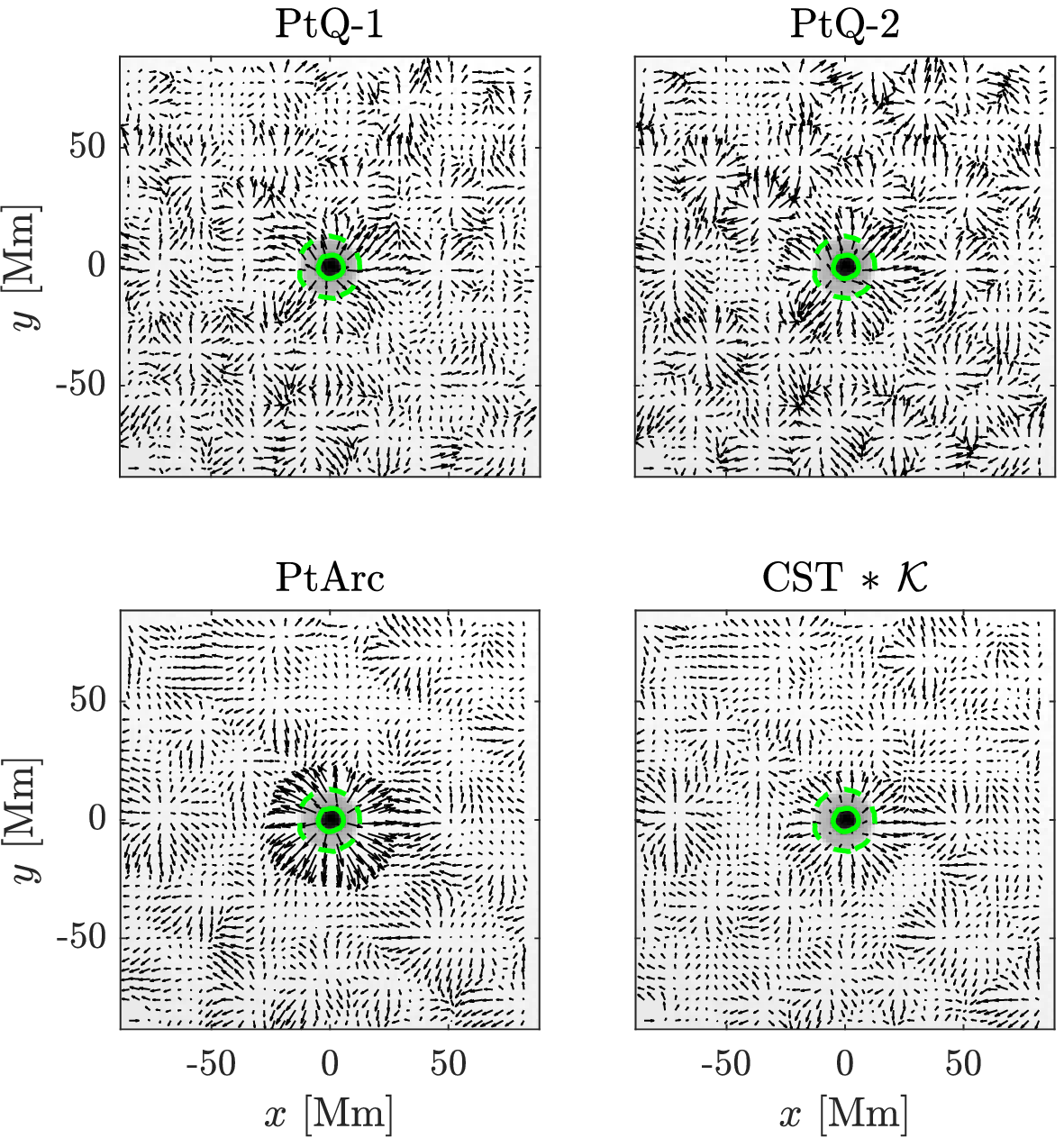}
{Inversions for horizontal flows around the active region NOAA~11084 on 1~July 2010. Top: Inversions in PtQ geometries. Bottom: Inversions in PtArc geometries and CST based flows. The reference arrows at the bottom left corners correspond to 250~\mps{} and all have the same length. The full and the dashed contours correspond to umbra and penumbra. The background images are the associated intensitygrams. }
{fig:inversion_check_spot_appendix}

\wfigure[!ht]{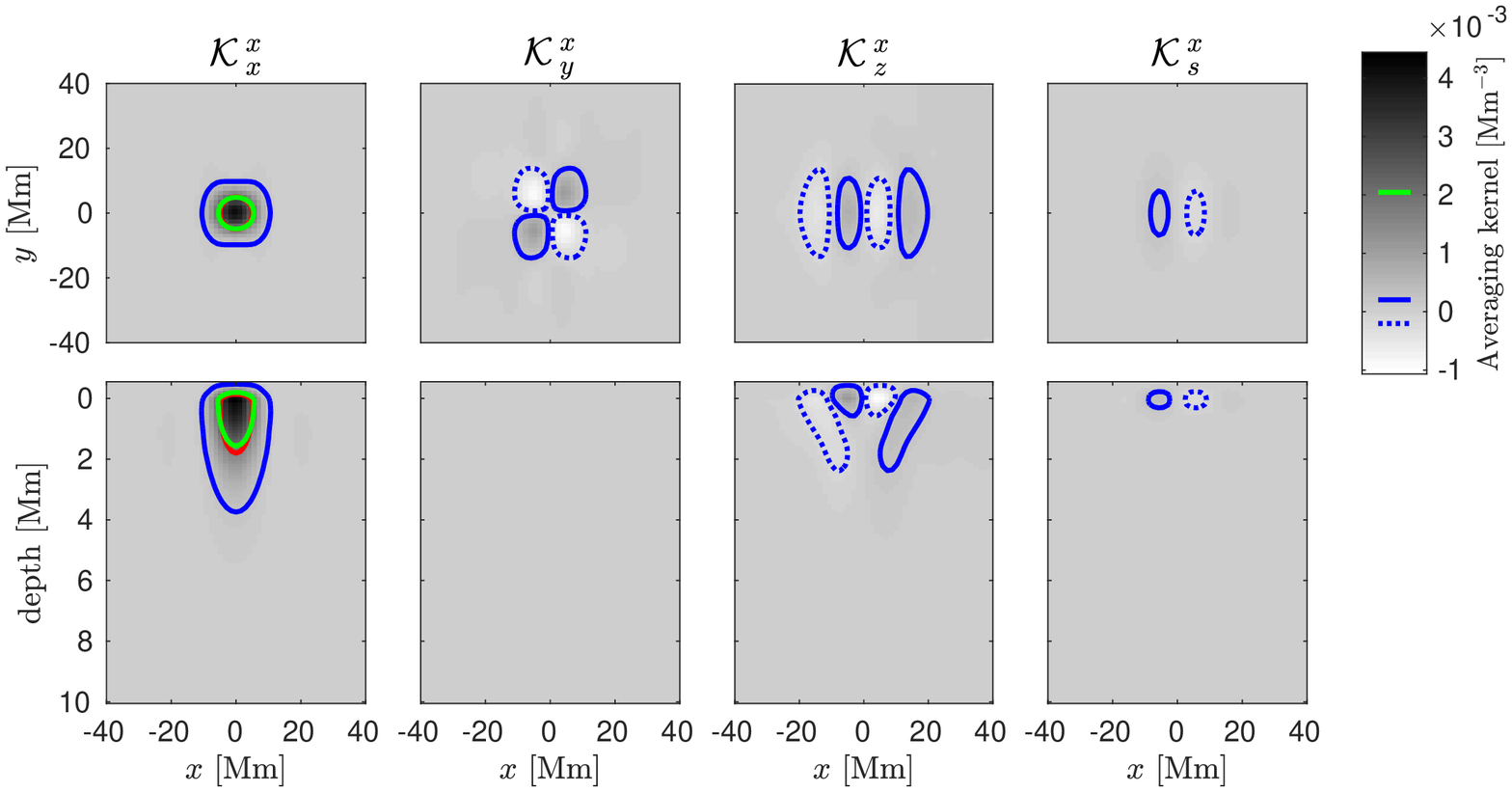}
{Averaging kernel for the PtQ-1 inversion for the surface longitudinal flow. Top row: Surface cuts. Bottom row: Vertical cuts at $y = 0$. Columns: Localisations of longitudinal flows, latitudinal flows, vertical flows, and \ssp{} (from left to right). The red and the green contours correspond to the half-maximum of the target function and the averaging kernel, respectively. The full and the dotted blue contours correspond to $+5\%$ and $-5\%$ of the maximum of the averaging kernel. }
{fig:rakern_appendix}

In order to minimise the cross-talk between horizontal flows, we performed another set of PtQ inversion (hereafter referred to as PtQ-2). In these inversions, we combined both horizontal-flow components and e-w and n-w averaging geometries in one inversion setup; therefore, the cross-talk can be suppressed by the technique introduced by \citet{Svanda_2011} that is an inseparable part of our inversion pipelines. The averaging kernel for $v_x$ component is shown in Fig.~\ref{fig:rakern2_appendix}. We note that $\mathrm{S/N} = 7.8$ in this case. 

\wfigure[!ht]{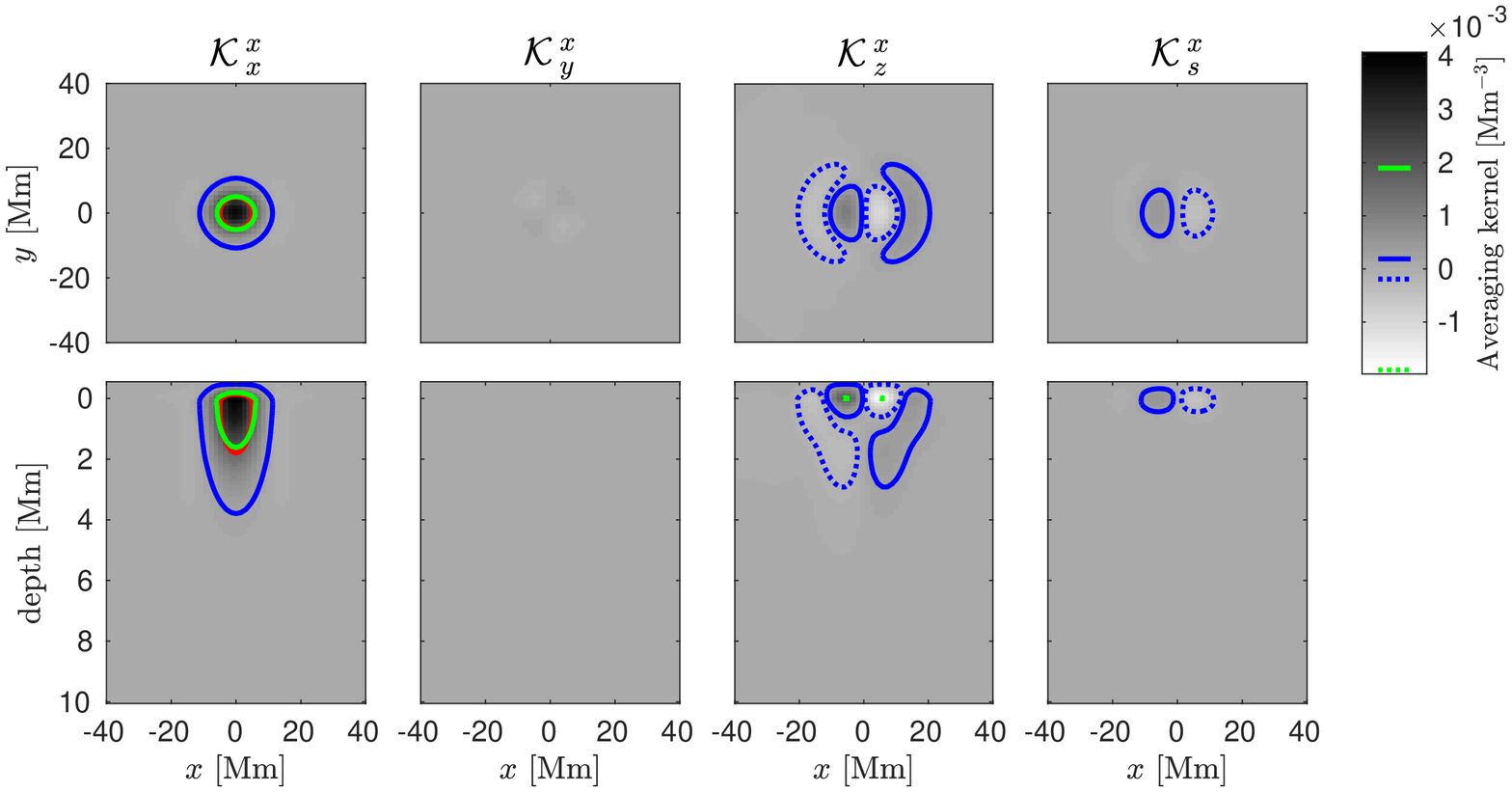}
{Averaging kernel for the PtQ-2 inversion for the surface longitudinal flow. Top row: Surface cuts. Bottom row: Vertical cuts at $y = 0$. Columns: Localisations of longitudinal flows, latitudinal flows, vertical flows, and \ssp{} (from left to right). The red and the green full and dotted contours correspond to the half-maximum of the target function and the half-maximum and the minus half-maximum the of averaging kernel, respectively. The full and the dotted blue contours correspond to $+5\%$ and $-5\%$ of the maximum of the averaging kernel. We note that the cross-talk contribution from the perpendicular horizontal component (second column) is greatly minimised as compared to Fig.~\ref{fig:rakern_appendix}.}
{fig:rakern2_appendix}

The correlation coefficients between PtQ-1 and $\mathrm{CST} \ast \mathcal{K}$ in the same format as in Tables~\ref{tab:summarise_vx} and \ref{tab:summarise_vy} are 0.81 (0.69) for $v_x$ and 0.74 (0.68) for $v_y,$ and for the PtQ-2 numbers these are 0.87 (0.80) for $v_x$ and 0.80 (0.76) for $v_y$. These are lower than correlation coefficients between PtArc and $\mathrm{CST} \ast \mathcal{K}$ presented in Tables~\ref{tab:summarise_vx} and \ref{tab:summarise_vy}. The slopes $s$ corresponding to the values in parentheses in Tables~\ref{tab:summarise_vx} and \ref{tab:summarise_vy} between PtQ-1, PtQ-2, and $\mathrm{CST} \ast \mathcal{K}$ are 0.88 for PtQ-1 (both components) and 1.17 (PtQ-2, $v_x$) and 1.18 (PtQ-2, $v_y$). All these values are farther from unity than the corresponding values between PtArc and $\mathrm{CST} \ast \mathcal{K}$. The spatial extent of the moat still seems larger in the case of the PtQ models as seen in Fig.~\ref{fig:radial_profile_vh_appendix}. The PtQ-1 and the PtQ-2 moat flows are about 25\% and 10\% more extended in comparison with PtArc and CST, respectively. 

In the moat region, the PtQ-derived horizontal flow magnitude is about a factor of 2 smaller than the magnitude of the flows inverted using PtA and PtArc approaches. This magnitude decrease is probably due to the affection of the travel times by the presence of the magnetic field nearby. In the travel-time maps, the east--west and north--south travel times are greatly affected in the region of the sunspot, which inevitably leads to the suppression of the derived flow magnitude. On the other hand, the out--in travel time has a very strong magnitude in the sunspot region (see Fig.~\ref{fig:tt_arc_vs_ann}; leaving it proper interpretation aside), which explains why the PtA inversion (which combined not only the directional e-w and n-s but also the omnidirectional o-i geometry) yields a much larger flow amplitude. Therefore, we must conclude that even the very simple PtQ inversions are affected by the presence of the magnetic field in the sunspot in a much larger region, and as a consequence the topology of the flows is different from the PtArc one-sided inversions.

\wfigure{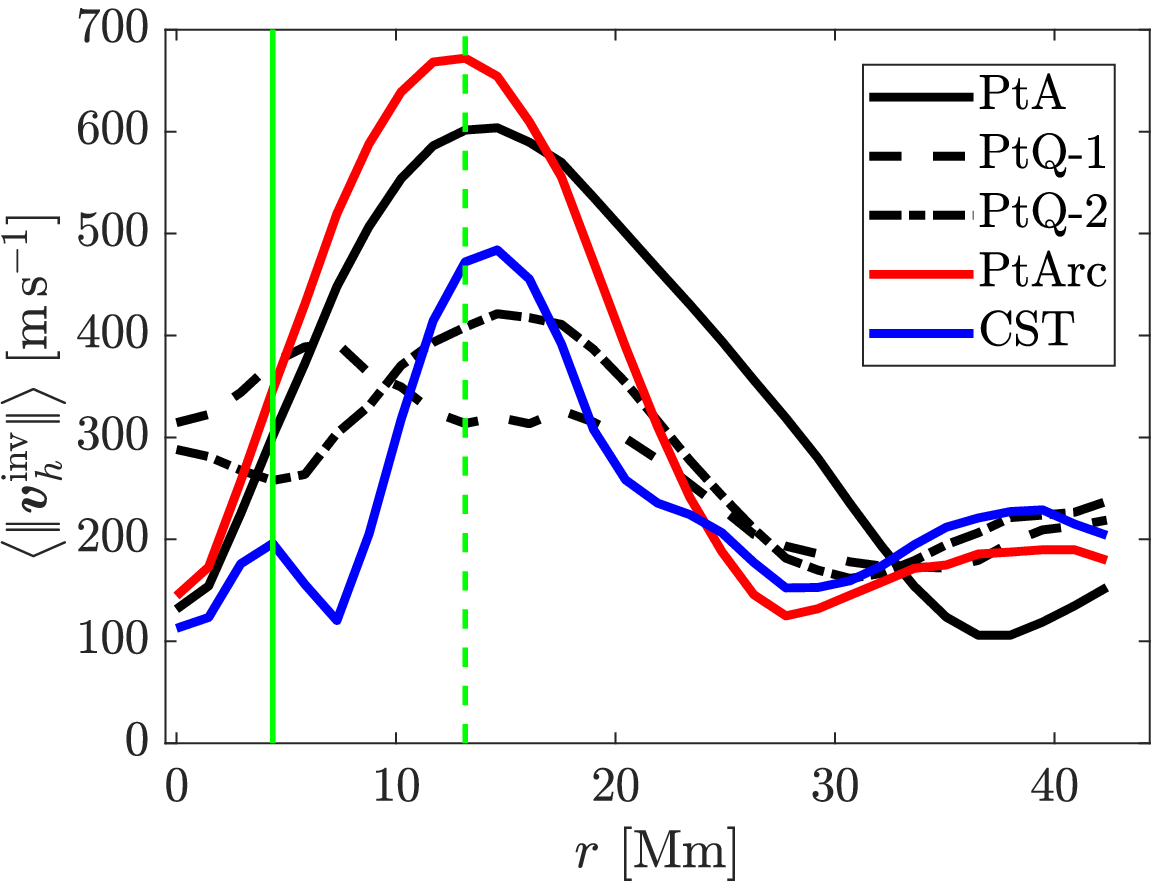}
{Radial profiles of azimuthally averaged speeds. The local minima of the speed are at distances 37.3, 34.6, 30.9, 28.1, and 28.5~Mm for PtA, PtQ-1, PtQ-2, PtArc, and CST respectively. The green vertical lines indicate the edges of umbra and penumbra.}
{fig:radial_profile_vh_appendix}

\end{document}